\documentclass[pdflatex,sn-mathphys-num]{sn-jnl}% Math and Physical Sciences Numbered Reference Style

\usepackage{graphicx}%
\usepackage{multirow}%
\usepackage{amsmath,amssymb,amsfonts}%
\usepackage{amsthm}%
\usepackage{mathrsfs}%
\usepackage[title]{appendix}%
\usepackage{xcolor}%
\usepackage{textcomp}%
\usepackage{manyfoot}%
\usepackage{booktabs}%
\usepackage{algorithm}%
\usepackage{algorithmicx}%
\usepackage{algpseudocode}%
\usepackage{listings}%

\theoremstyle{thmstyleone}%
\theoremstyle{thmstyletwo}%
\theoremstyle{thmstylethree}%

\begin{document}

\title[Article Title]{Noise-Assisted Quantum Entanglement: Memory-Enhanced Coherence in Stochastic Jaynes-Cummings Dynamics}

\author[1]{\fnm{Fang} \sur{Wang}}\email{wfaugust@gmail.com}
\author[1]{\fnm{Quanzhen} \sur{Ding}}\email{xyzdqz@gmail.com}
\author[2]{\fnm{Jun} \sur{Jing}}\email{jingjun@zju.edu.cn}
\author*[1]{\fnm{Ting} \sur{Yu}}\email{tyu1@stevens.edu}

\affil*[1]{\orgdiv{Department of Physics}, \orgname{Stevens Institute of Technology}, \orgaddress{\street{1 Castle Point Terrace}, \city{Hoboken}, \postcode{07310}, \state{New Jersey}, \country{USA}}}
\affil[2]{\orgdiv{School of Physics}, \orgname{Zhejiang University}, \orgaddress{\street{866 Yuhangtang Rd}, \city{Hangzhou}, \postcode{310027}, \state{Zhejiang}, \country{China}}}

%%=============================================================%%
%% ABSTRACT
%% [R3-a] Removed all mathematical expressions from abstract.
%%=============================================================%%
\abstract{We investigate qubit entanglement dynamics in a two-qubit cavity
system with stochastically fluctuating coupling strength, comparing white noise
(Markovian) and Ornstein-Uhlenbeck noise (non-Markovian) processes.
We analytically solve the system dynamics for both Fock and coherent cavity
states and systematically characterize how initial qubit configurations affect
entanglement generation and preservation.
Our results reveal distinct dynamical regimes: in the Fock-state cavity,
initial states in the single-excitation subspace generate robust qubit
entanglement in decoherence-free subspaces, while initially entangled states
in the double-excitation subspace exhibit inevitable decay without protective
symmetries.
For coherent cavities, we identify a critical photon-number threshold in the
intermediate coherent-state amplitude regime where dynamics transition from
low-frequency to high-frequency entanglement oscillations, while universal
entanglement loss occurs at large amplitudes.
Remarkable resilience against entanglement sudden death emerges under strong
non-Markovian noise with long correlation times, where repeated revivals occur.
These findings establish cavity photon statistics and environmental memory time
as key control parameters for engineering qubit entanglement dynamics in
qubit-cavity systems, revealing how system-environment correlations govern
quantum information preservation.}

\keywords{Jaynes-Cumming Model · Open quantum system · Entanglement sudden death · Non-Markovian dynamics}

\maketitle

\section{INTRODUCTION}
\label{sec:intro}

The Jaynes-Cummings model (JC model), introduced by Jaynes and Cummings in 1963\cite{jaynes1963comparison}, stands as a cornerstone in quantum optics and quantum field theory. This fundamental model describes a two-level atom (qubit) interacting with a single quantized optical cavity mode, where the cavity acts as an electromagnetic radiation bath that induces spontaneous emission and absorption processes. Let $\omega$ denote the atomic transition frequency and $\omega_f$ denote the cavity mode frequency. Under the resonance condition $\omega_f = \omega$, the standard JC Hamiltonian in the interaction picture takes the form:
\begin{equation}
H_{JC} = g(a^\dagger \sigma_- + a \sigma_+),
\end{equation}
where $g$ is the coupling strength, $a^\dagger$ ($a$) represents the photon creation (annihilation) operator, and $\sigma_\pm$ are the atomic raising and lowering operators. The JC model has provided the theoretical foundation for understanding critical phenomena, including light-matter interactions, cavity quantum electrodynamics (QED), and circuit QED\cite{yoshie2004vacuum, deppe2008two, PhysRevLett.110.193602, PhysRevLett.112.013601}.

Bipartite entanglement of a quantum open system can exhibit a finite-time, non-asymptotic collapse under environmental decoherence\cite{PhysRevLett.101.080503, yu2009sudden, yonacc2006sudden, yu2006sudden, eberly2007end, almeida2007environment, PhysRevLett.93.140404, PhysRevLett.97.140403, PhysRevA.73.040305}. The study of these phenomena has provided some useful insights into the more general temporal evolution of entanglement in dissipative light--matter platforms, revealing regimes in which quantum correlations may decay exponentially, collapse at a well-defined time, or later revive when memory or structured reservoirs are present. Understanding the temporal evolution of entanglement is essential for quantum technologies, since the viability of quantum resources depends on their coherence time and correlation lifetime, which determine the operational window for quantum computing, cryptography, and quantum communication protocols\cite{zhang2024entanglement, zanardi2021information, sharma2021quantum, gour2021entanglement, wu2023entanglement, guo2022genuine, sauer2021entanglement, yang2024characterizing, li2024platform}. Moreover, the ability to predict, preserve, or restore entanglement dynamics in realistic noisy environments therefore has direct implications for the durability and controllability of quantum optical information carriers\cite{gebhart2023learning, han2021experimental, guimaraes2023noise, white2022characterization, long2022entanglement, dalzell2024random, youssry2020characterization}. Extensive research has investigated temporal behaviors of entanglement including decay, revival, and stabilization within various extensions of the standard JC framework. The Tavis--Cummings model has been instrumental in understanding collective entanglement dynamics and collapse--revival behavior\cite{PhysRev.170.379, PhysRevA.68.062316}, while studies have explored entanglement dynamics in driven systems under classical noise\cite{rodriguez2013classical, PhysRevA.45.5113, PhysRevA.62.043812}, thermalized environments\cite{PhysRevA.65.040101, yonacc2008qubit, yonacc2006sudden}, and cavity-mediated entanglement transfer between spatially separated qubits\cite{zhou2006entanglement}. Investigations of multiple coupled JC systems have revealed that disorder in atom--cavity interactions can mitigate entanglement decay and delay correlation loss\cite{ghoshal2020population, qiang2018concurrence}, while controlled photon-hopping between cavities can completely preclude sudden entanglement loss by redistributing excitations\cite{pandit2018effects}. Recent work on non-Markovian dynamics\cite{PhysRevA.81.062124, PhysRevLett.105.240403, PhysRevA.84.032101, chen2016non} has shown that non-Markovianity can enhance entanglement through memory-assisted correlation revival and information backflow\cite{motavallibashi2021non}, and quantum discord remains robust even when entanglement temporarily vanishes\cite{mirmasoudi2018dynamics, ban2022quantum, castanos2024coherent}. Despite this progress, limited investigation exists for scenarios where multiple qubits interact through a shared cavity while the coupling parameters fluctuate stochastically with a general correlation function. This configuration is particularly relevant for emerging quantum photonic architectures and cavity QED networks, where understanding entanglement decay, revival, and correlation stabilization under memory-bearing stochastic coupling noise remains a timely open challenge.

In this paper, we analyze a novel JC model comprising two qubits coupled to a shared optical resonator with stochastic coupling strength. The system incorporates coupling coefficients following white noise (Markovian) or OU noise (non-Markovian) processes, enabling systematic investigation of dynamical regime transitions. We provide analytical solutions and numerical simulations for a cavity prepared either in Fock or coherent state, quantifying qubit-qubit entanglement dynamics through concurrence measurements~\cite{wootters2001entanglement, yu2005evolution}. Our analysis reveals that entanglement between the two qubits depends critically on initial states, particularly in low-photon regimes and for certain initial probability distributions of the cavity mode. While certain decoherence-free configurations preserve qubit-qubit quantum correlations indefinitely, other initial states exhibit inevitable decoherence, highlighting the fragility of pre-existing two-qubit entanglement in noisy environments. This asymmetry indicates that maintaining quantum correlations between qubits requires careful initial state engineering rather than relying solely on dynamical protection. Higher cavity excitations suppress qubit-qubit entanglement generation and accelerate decoherence, making the vacuum or low-photon regime optimal for two-qubit entanglement protocols. Non-Markovian environments with extended correlation times can outperform traditional Markovian noise through memory-assisted revival mechanisms. For coherent cavity states, we identify distinct optimal operating regimes: low amplitudes maximize robustness under noise-free conditions, while intermediate amplitudes enable qubit-qubit entanglement revival in engineered non-Markovian environments. High amplitudes should be avoided due to two-qubit entanglement death. These results challenge conventional assumptions about environmental decoherence and demonstrate that carefully structured reservoirs with slower fluctuations can protect quantum coherence, opening possibilities for noise-assisted quantum phenomena and providing practical guidance for optimizing two-qubit entanglement-based protocols in realistic experimental implementations.

%%=============================================================%%
%% [R2-1] Added novelty paragraph before "The remainder..."
%%=============================================================%%
Compared with existing literature, the principal novelties of the present work are threefold.
First, we treat a two-qubit system sharing a single cavity mode with a
stochastic coupling whose correlation function interpolates continuously
between the Markovian and strongly non-Markovian limits via the OU
parameter $\gamma$; most prior analytical studies address single-qubit JC
models or restrict to specific limiting noise cases.
Second, we derive closed-form analytical expressions for the noise-averaged
concurrence for both Fock and coherent cavity states and for two physically
distinct classes of initial two-qubit state, enabling systematic
parameter-space exploration without Monte Carlo sampling.
Third, we identify a photon-number threshold in the intermediate
coherent-state amplitude regime that separates revival-supporting from
revival-free dynamics, providing concrete design criteria for entanglement
protocols in cavity-QED and circuit-QED platforms.

The remainder of this paper is organized as follows: Section II introduces our qubit-resonator model and presents the explicit Hamiltonian formulation. Sections III and IV provide analytical solutions for the time evolution and present numerical simulations of entanglement dynamics measured through concurrence for the Fock state cavity and the coherent state cavity, respectively. Section V discusses conclusions and implications. The appendices provide detailed explanations: Appendix~\ref{app:note_fock_special_case} presents the special cases for the Fock state cavity when $n=0$ and $n=1$; Appendix~\ref{app:coherent2} provides the derivation for the time evolution of the coherent cavity case with the initial state $\left(\cos\theta\,|00\rangle + \sin\theta\,|11\rangle\right)\otimes|\alpha\rangle$; and Appendix~\ref{app:Scaling} discusses the scaling of cavity frequency and coupling fluctuations.

\section{MODELS AND HAMILTONIAN}

\subsection{Model}

\begin{figure}
\centering
\includegraphics[width=0.85\columnwidth, clip=true]{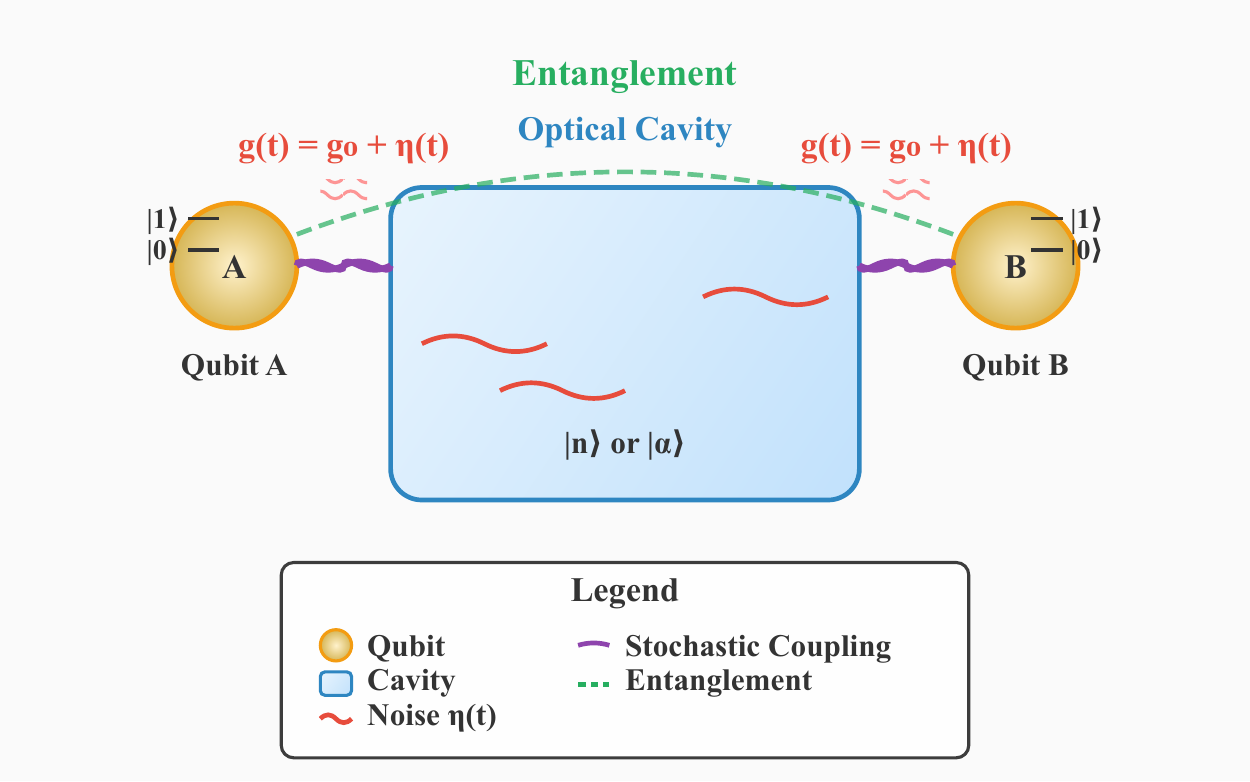}
\vspace{2mm}
\caption{Schematic of the JC model with stochastic coupling $g(t) = g_0 + \eta(t)$ between two qubits (A and B) and an optical cavity. The stochastic noise $\eta(t)$ (white noise or OU noise) affects the qubit-cavity coupling strength, generating entanglement between the qubits through cavity-mediated interactions.}
\vspace{0mm}
\label{fig:00}
\end{figure}

We consider a system consisting of two qubits interacting linearly with a single-mode cavity field. The total Hamiltonian in the Schr\"{o}dinger picture can be written as
\begin{small}
\begin{equation}
H_{\mathrm{tot}}(t) = \frac{\omega}{2} \left( \sigma_{z}^{A} + \sigma_{z}^{B} \right) + \omega_{f} a^{\dagger} a + g(t) \left[ a^{\dagger} \left( \sigma_{-}^{A} + \sigma_{-}^{B} \right) + a \left( \sigma_{+}^{A} + \sigma_{+}^{B} \right) \right]
\label{eq:total_hamiltonian}
\end{equation}
\end{small}
where qubits \(A\) and \(B\) have identical transition frequency \(\omega\), and \(\sigma_z^{j}\) (\(j = A,B\)) denotes the Pauli-\(z\) operator for qubit \(j\). The operators \(\sigma_-^{j}\) and \(\sigma_+^{j}\) are, respectively, the lowering and raising operators for qubit \(j\). The operators \(a^\dagger\) and \(a\) are the bosonic creation and annihilation operators of the cavity mode, which has frequency \(\omega_f\). The coupling strength between each qubit and the cavity is time-dependent and denoted by \(g(t)\), accounting for possible fluctuations or modulation. In this work, we focus on the resonant case $\omega_f = \omega$, where the qubits and the cavity mode are frequency-matched, allowing for maximal energy exchange. Throughout this work, we adopt natural units where $\hbar = 1$ and set $\omega = 1$, so that all energies, frequencies, and the coupling $g_0$ are measured in units of $\omega$; in particular $g_0/\omega = g_0$ numerically.

%%=============================================================%%
%% [R1-1a, R1-1b] Corrected description of interaction picture
%% transformation and removed erroneous RWA statement.
%%=============================================================%%
Moving to the interaction picture with respect to the free Hamiltonian
\[
H_0 = \frac{\omega}{2} \left( \sigma_{z}^{A} + \sigma_{z}^{B} \right) + \omega a^{\dagger} a,
\]
the total Hamiltonian reduces to
\begin{equation}
H_{\mathrm{tot}}^{I}(t) = g(t) \left[ a^{\dagger} \left( \sigma_{-}^{A} + \sigma_{-}^{B} \right) + a \left( \sigma_{+}^{A} + \sigma_{+}^{B} \right) \right].
\label{eq:interaction_hamiltonian}
\end{equation}
Under the resonance condition $\omega_f = \omega$, the free Hamiltonian $H_0$ commutes with the total excitation number $\hat{N} = a^\dagger a + \frac{1}{2}(\sigma_z^A + \sigma_z^B) + 1$. The transformation to the interaction picture therefore removes only the free-energy phase factors without altering the operator structure of the coupling term. We note that the Jaynes-Cummings Hamiltonian in Eq.~\eqref{eq:total_hamiltonian} is itself obtained from the full atom--field dipole interaction (Dicke model) by a prior application of the rotating-wave approximation, which is valid when $g_0 \ll \omega$; no further application of the RWA is required or performed in what follows.

The time-evolution operator in the interaction picture is
\begin{equation}
U(t) = \mathcal{T} \exp\left[ -i \int_0^t H_{\mathrm{tot}}^I(\tau) \, d\tau \right],
\label{eq:evolution_operator}
\end{equation}
where \(\mathcal{T}\) denotes the time-ordering operator. If the interaction Hamiltonian is self-commuting at different times, i.e.,
\[
[H_{\mathrm{tot}}^I(t), H_{\mathrm{tot}}^I(s)] = 0 \quad \forall \, t, s,
\]
which holds since \(g(t)\) is a scalar function and the operator structure of \(H_{\mathrm{tot}}^I\) is time-independent, then the time-ordering symbol can be dropped.

\subsection{Stochastic Noise Coupling}

The coupling factor $g(t) = g_0 + \eta(t)$ consists of a constant term $g_0$ and a fluctuating term $\eta(t)$. In realistic experimental implementations, the coupling strength between qubits and cavity mode is subject to various sources of fluctuations that make it inherently stochastic\cite{PhysRevA.69.052115, PhysRevA.84.032101}. These fluctuations arise from various physical mechanisms including cavity parameter variations due to thermal expansion, mechanical vibrations, or electromagnetic field fluctuations that modify the cavity geometry and mode structure\cite{PhysRevLett.105.240403}, laser intensity and phase noise in driven systems where pump laser instabilities directly translate into coupling strength variations\cite{PhysRevA.82.062114}, atomic motion and positioning uncertainties in systems with mobile qubits leading to spatial variations in the coupling strength\cite{PhysRevB.68.165322}, and environmental electromagnetic interference from external sources that can modulate the effective interaction strength\cite{breuer2002theory}.

%%=============================================================%%
%% [R2-2] Clarified classical vs. quantum nature of the noise.
%%=============================================================%%
We emphasize that $\eta(t)$ represents a \emph{classical} stochastic process:
the coupling fluctuations are modeled as a classical random variable with
prescribed statistics, rather than as an operator-valued quantum bath.
This description is appropriate when the environmental degrees of freedom
(thermal phonons, laser phase noise, cavity mechanical vibrations) are in a
high-temperature or large-occupation regime well approximated by classical
statistics.
For a purely quantum environment one would instead derive a Lindblad or
Nakajima--Zwanzig master equation by tracing over the bath; the exponential
OU correlation function then arises naturally from a Lorentzian bath spectral
density.
In either formulation the functional form of the decoherence factor (see
$A_n$ in Eq.~\eqref{eq:An} below) is unchanged at the level of second-order
cumulants, which justifies our classical treatment as a physically motivated
and analytically tractable approximation.

The choice between white noise and OU noise models reflects different physical scenarios: white noise corresponds to rapidly fluctuating environments with negligible correlation times compared to the system dynamics, while OU noise captures environments with finite memory effects, such as those arising from mechanical resonances or filtered electromagnetic noise sources\cite{PhysRevA.90.042108, chen2016non}. This stochastic modeling thus provides a realistic framework for understanding cavity QED experiments under practical conditions\cite{RevModPhys.88.021002}.

The noise term $\eta(t)$ has a mean value and correlation function given by
\begin{small}
\begin{align}
\mathcal{M}[\eta(t)] &= 0, \nonumber \\
\mathcal{M}[\eta(t) \eta(s)] &= \frac{\Gamma \gamma}{2} e^{-\gamma |t - s|}.
\label{eq:ou_correlation}
\end{align}
\end{small}
where $\Gamma$ is the decay rate and $\gamma$ is the memory time parameter. Here, $\mathcal{M}[\cdot]$ denotes the ensemble average over the noise. In the limit $\gamma \rightarrow \infty$, the OU noise reduces to white noise, and the correlation function becomes
\begin{equation}
\alpha(t, s) = \Gamma \delta(t - s),
\label{eq:white_correlation}
\end{equation}
which corresponds to Markovian dynamics.

Beyond this phenomenological description, small fluctuations of the cavity size provide a concrete microscopic mechanism that induces correlated fluctuations of both the cavity frequency $\omega_f$ and the coupling strength $g(t)$. For a Fabry--Perot cavity of length $L$, the standing-wave mode frequency scales as $\omega_f(L) = m\pi c/(nL)$, so that a small length change $\delta L$ produces
\begin{equation}
\frac{\delta \omega_f}{\omega_f} \approx -\frac{\delta L}{L}.
\end{equation}
Similarly, the vacuum Rabi coupling $g \propto \sqrt{\omega_f/V}$ with mode volume $V \sim A L$ leads to $g \propto 1/L$ in this simple geometry, and hence
\begin{equation}
\frac{\delta g}{g} \approx -\frac{\delta L}{L}.
\end{equation}
If the length change arises from thermal expansion, $\delta L \approx \alpha_L L \Delta T$, the relative fluctuations of $\omega_f$ and $g$ scale linearly with $\alpha_L \Delta T$. When $\delta L(t)$ follows an OU process, both $\omega_f(t)$ and $g(t)$ inherit OU statistics with finite correlation time, providing a physically grounded realization of the stochastic coupling model used in this work. The detailed description can be found in Appendix~\ref{app:Scaling}.

%%=============================================================%%
%% [R1-2] Justified weak/strong coupling via dimensionless g0/omega.
%%=============================================================%%
In the numerical results that follow, we compare two representative values of
the dimensionless coupling $g_0/\omega$, which characterises the interaction
strength between the qubits and the cavity.
Throughout this work we set $\omega = 1$ in natural units, so that
$g_0/\omega = g_0$ numerically.
With this convention, $g_0 = 0.5$ gives $g_0/\omega = 0.5 \ll 1$,
placing the system well within the weak-coupling regime where the JC
Hamiltonian is valid; we refer to this as the \emph{weak coupling} case.
Conversely, $g_0 = 5$ gives $g_0/\omega = 5 \gg 1$; while this formally
lies outside the strict perturbative regime of the standard JC model,
it provides a useful strong-coupling limit for exploring non-perturbative
dynamics, and physical implementations in this regime are accessible in
the ultrastrong-coupling regime of circuit QED\cite{RevModPhys.88.021002}.
We refer to this as the \emph{strong coupling} case.

In the following sections, we analyze the system dynamics under different initial conditions, focusing on the effects of noise on qubit entanglement behavior and the transition between Markovian and non-Markovian regimes.

\section{FOCK STATE CAVITY}
\subsection{Case I: Initial State $\cos\theta\,|01\rangle + \sin\theta\,|10\rangle$}

First, we consider the initial total state in the form
\begin{equation}
\begin{aligned}
|\psi\rangle=\left(\cos\theta|01\rangle+\sin\theta|10\rangle \right) \otimes |n\rangle,
\end{aligned}
\end{equation}
where the two qubits are initially coupled with a cavity which is prepared in a Fock state $|n\rangle$.

The total system evolves in the Hilbert space spanned by the following four states: $|1\rangle_{n}=|0,0,n+1\rangle$, $|2\rangle_{n}=|0,1,n\rangle,$ $|3\rangle_{n}=|1,0,n\rangle,$ $|4\rangle_{n}=|1,1,n-1\rangle$. For simplicity, we ignore the $n=0$ and $n=1$ case in the main text and leave it in the appendix. Under the interaction picture, the commutation relation $[H_{\mathrm{tot}}^{\mathrm{I}}(t), H_{\mathrm{tot}}^{\mathrm{I}}(s)]=0$ holds, allowing us to calculate the time evolution propagator directly as:
\begin{equation}
G_{\mathrm{tot}}(t) = \mathcal{T} \exp\left[ -i \int_0^t H_{\mathrm{tot}}^{\mathrm{I}}(\tau) \, d\tau \right].
\label{eq:propagator}
\end{equation}

Our objective is to determine how entanglement between the two qubits evolves from this initial state through their mutual interaction with the cavity mode. The initial density matrix of the total system in the basis defined above is
\begin{equation}
\rho_{\mathrm{tot}}(0)=\left(\begin{array}{cccc}
0 & 0 & 0 & 0\\
0 & \cos^{2}\theta & \cos\theta \sin\theta & 0\\
0 & \cos\theta \sin\theta & \sin^{2}\theta & 0\\
0 & 0 & 0 & 0
\end{array}\right).
\end{equation}

\begin{figure}
    \centering
    \includegraphics[width=1.0\columnwidth, clip=true]{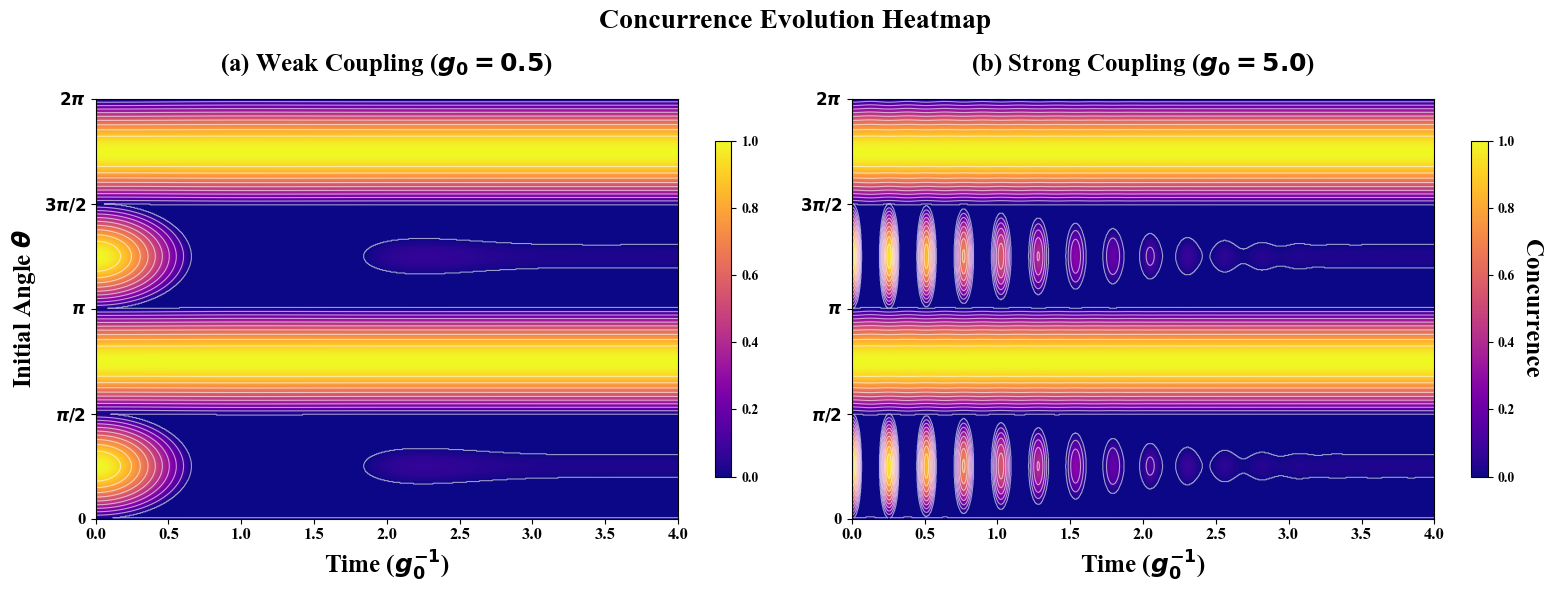}
    \vspace{-3mm}
    %%=============================================================%%
    %% [R3-f] Shortened figure caption; interpretive content moved to body.
    %%=============================================================%%
    \caption{Concurrence evolution for initial states $(\cos\theta|01\rangle+\sin\theta|10\rangle) \otimes |1\rangle$ under OU noise ($\Gamma=1$, $\gamma=0.1$): (a) weak coupling regime ($g_0=0.5$, $g_0/\omega=0.5$); (b) strong coupling regime ($g_0=5$, $g_0/\omega=5$).}
    \vspace{3mm}
    \label{fig:Fock_separable_state_concurrence}
\end{figure}

After time evolution, the density matrix becomes
\begin{equation}
\label{eq:total_density_evolution}
\rho_{\mathrm{tot}}(t)=\mathcal{M}\left[G_{\mathrm{tot}}(t)\rho_{\mathrm{tot}}(0)G_{\mathrm{tot}}^{\dagger}(t) \right],
\end{equation}
where $\mathcal{M}[\cdot]$ denotes the stochastic average over noise realizations. To examine the entanglement dynamics between the two qubits, we trace out the cavity degrees of freedom to obtain the reduced density matrix:
\small{
\begin{eqnarray}
\rho(t)=\mathcal{M}\left[\left(\begin{array}{cccc}
S_{1,n}^{2} & 0 & 0 & 0\\
0 & X_{1,n}^{2} & X_{1,n}X_{2,n} & 0\\
0 & X_{1,n}X_{2,n} & X_{2,n}^{2} & 0\\
0 & 0 & 0 & S_{2,n}^{2}
\end{array}\right)\right], \label{special1}
\end{eqnarray}
}
with
\footnotesize{
\begin{eqnarray*}
    X_{1,n}&=&-\frac{\sqrt{2}}{2} \sin \left(\theta - \frac{\pi}{4}\right) + \frac{\sqrt{2}}{2}\cos x_{n}\cos \left(\theta - \frac{\pi}{4}\right),\\
    X_{2,n}&=&\frac{\sqrt{2}}{2} \sin \left(\theta - \frac{\pi}{4}\right) + \frac{\sqrt{2}}{2}\cos x_{n}\cos \left(\theta - \frac{\pi}{4}\right),\\
    S_{1,n}&=&\sqrt{\frac{n+1}{2n+1}}\sin x_{n}\cos \left(\theta - \frac{\pi}{4}\right),\\
    S_{2,n}&=&\sqrt{\frac{n}{2n+1}}\sin x_{n}\cos \left(\theta - \frac{\pi}{4}\right),
\end{eqnarray*}
}
\normalsize
where the phase accumulation is given by
\begin{equation}
x_n = \sqrt{2(2n+1)}\Omega(t),
\label{eq:phase_accumulation}
\end{equation}
with $\Omega(t) = \int_0^t g(\tau)d\tau$.

To quantify the entanglement between the two qubits, we employ Wootters' concurrence~\cite{wootters2001entanglement}. Concurrence $C$ is defined as
\begin{equation}
C(\rho)=\max\{0, \sqrt{\lambda_{1}}-\sqrt{\lambda_{2}}-\sqrt{\lambda_{3}}-\sqrt{\lambda_{4}}\},
\end{equation}
where $\lambda_{i}$, arranged in decreasing order, are the eigenvalues of the auxiliary matrix $\zeta=\rho\tilde{\rho}$ with
\begin{equation}
\tilde{\rho}=(\sigma_{y}\otimes\sigma_{y})\rho^{\star}(\sigma_{y}\otimes\sigma_{y}).
\end{equation}
Here, $\rho^{\star}$ denotes the complex conjugation of $\rho$ in the standard basis and $\sigma_{y}$ is the Pauli Y matrix in the same basis.
%%=============================================================%%
%% [R3-b] Corrected/clarified the C=1 statement.
%%=============================================================%%
Concurrence provides a convenient normalized measure of entanglement:
$0\leq C(\rho)\leq 1$, where $C(\rho)=0$ indicates a separable (unentangled)
state, and $C(\rho)=1$ if and only if $\rho$ is a pure maximally entangled
(Bell) state.

Since the reduced density matrix for the two-qubit system always has the $X$-form~\cite{yu2005evolution, yu2010entanglement}, we can directly calculate the concurrence as
\footnotesize{
\begin{eqnarray}
 C(\rho) &=& 2\max \{\left|\mathcal{M}\left[X_{1,n}X_{2,n}\right]\right|-\sqrt{\mathcal{M}\left[S_{1,n}^2\right] \mathcal{M}\left[S_{2,n}^2\right]} -\sqrt{\mathcal{M}\left[X_{1,n}^2\right] \mathcal{M}\left[X_{2,n}^2\right]}, 0 \} \nonumber\\
 &=& \max \{\left|\left(A_{n}+1\right) \cos^2 \left(\theta - \frac{\pi}{4}\right)- 1\right| -\frac{2\sqrt{n\left(n+1\right)}}{2n+1}\left(1-A_{n}\right)\cos^2 \left(\theta - \frac{\pi}{4}\right), 0 \},
 \end{eqnarray}
}
\normalsize
%%=============================================================%%
%% [R3-c] Added Gaussian moment identity justification before A_n.
%%=============================================================%%
where the noise-averaged quantity $A_n = \mathcal{M}[\cos^2 x_n]$ is evaluated
using the following procedure.
Since $\eta(t)$ is an OU process, $\Omega(t) = \int_0^t g(\tau)\,d\tau$ is a
Gaussian random variable with mean $\mathcal{M}[\Omega(t)] = g_0 t$ and
variance
\begin{equation}
\mathrm{Var}[\Omega(t)] = \Gamma t + \frac{\Gamma}{\gamma}\!\left(e^{-\gamma t} - 1\right).
\label{eq:var_Omega}
\end{equation}
All noise-averaged trigonometric functions are then evaluated using the
Gaussian moment identity
\begin{equation}
\mathcal{M}\!\left[e^{ikX}\right]
  = \exp\!\left(ik\,\mathcal{M}[X] - \tfrac{k^2}{2}\mathrm{Var}[X]\right),
\label{eq:gaussian_moment}
\end{equation}
which holds for any real constant $k$ and Gaussian variable $X$.
In particular, $\mathcal{M}[\cos x_n] = \mathrm{Re}\,\mathcal{M}[e^{ix_n}]$
and $\mathcal{M}[\cos^2 x_n] = \tfrac{1}{2}(1 + \mathrm{Re}\,\mathcal{M}[e^{2ix_n}])$
follow directly from Eq.~\eqref{eq:gaussian_moment}.
Applying these identities gives
\footnotesize{
 \begin{eqnarray}
A_{n} &=& \mathcal{M}\left[\cos^2 x_{n}\right]\nonumber\\
&=& \frac{1}{2}+\frac{1}{2}\Re\left\{\exp\left[\mathcal{M}\left[2ix_{n}\right] +\frac{1}{2}\mathrm{Var}\left[2ix_{n}\right]  \right]\right\}\nonumber\\
&=& \frac{1}{2}+\frac{1}{2}\cos\left(2\sqrt{4n+2}g_0t \right)\cdot\exp\left\{ -2(4n+2)\left[\Gamma t +\frac{\Gamma}{\gamma}(e^{-\gamma t}-1)\right]  \right\}.
\label{eq:An}
 \end{eqnarray}
}
\normalsize
%%=============================================================%%
%% [R2-3] Added explicit white-noise limit verification.
%%=============================================================%%
One may verify the Markovian limit explicitly: as $\gamma \to \infty$ with
$\Gamma$ fixed, the term $\frac{\Gamma}{\gamma}(e^{-\gamma t}-1) \to 0$, so
$A_n$ reduces to
\[
A_n\big|_{\gamma\to\infty}
= \frac{1}{2}+\frac{1}{2}\cos\!\left(2\sqrt{4n+2}\,g_0 t\right)
  e^{-2(4n+2)\Gamma t},
\]
recovering the white-noise decoherence result exactly.

This analytical expression reveals how the stochastic noise affects entanglement generation: the parameter $A_{n}$ encodes the combined effects of coherent evolution (through the cosine term) and noise-induced decoherence (through the exponential damping), determining whether and how quickly entanglement can be generated from the initial state.

\begin{figure}
    \centering
    \includegraphics[width=1.0\columnwidth, clip=true]{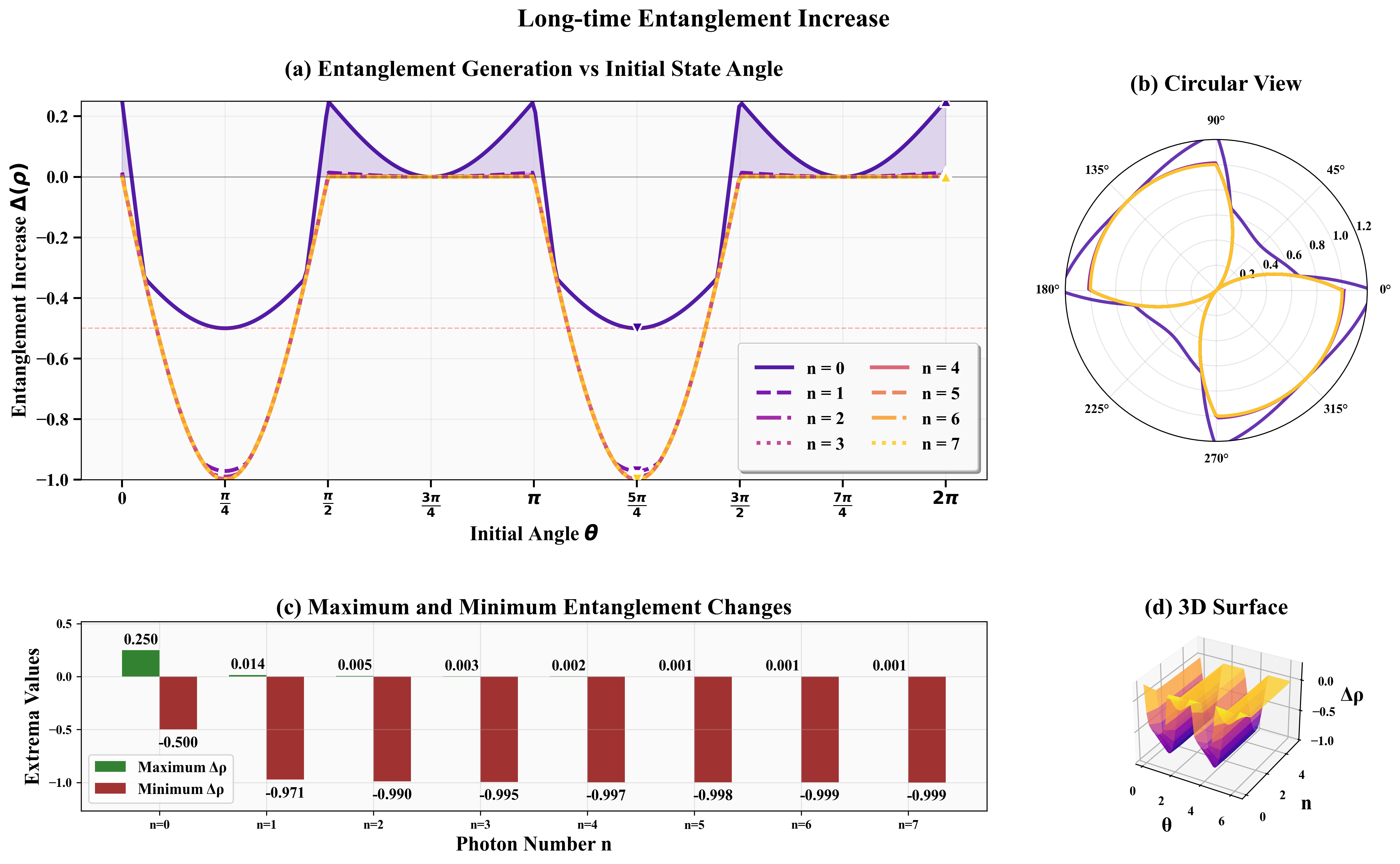}
    \vspace{-3mm}
    \caption{Long-time entanglement increase $\Delta(\rho)$: (a) versus initial angle $\theta$; (b) circular view; (c) maximum and minimum changes for different photon numbers $n$; (d) 3D surface plot. Parameters: $\Gamma=1$, $\gamma=0.1$.}
    \vspace{2mm}
    \label{fig:Fock_separable_state_long_time_entanglement_increase}
\end{figure}

Notably, the $n=0$ case also satisfies this equation for $C\left(\rho\right)$. Here, $A_{n}$ contains two terms: the first term is a constant $1/2$, and the second term is a product of a periodic function and a decaying function of time $t$. At time $t=0$, $A_{n}=1$. However, in the long-time limit when $t\rightarrow\infty$, $A_{n}\rightarrow 1/2$ and the system evolves into a steady state.

The concurrence evolution heatmaps in Fig.~\ref{fig:Fock_separable_state_concurrence} reveal distinct entanglement dynamics between the weak and strong coupling regimes. In the weak coupling case ($g_0=0.5$, $g_0/\omega=0.5$), we observe smooth, gradual evolution of entanglement with broad regions of sustained concurrence, particularly evident in the yellow bands around $\theta = \frac{\pi}{2}$ and $\theta = \frac{3\pi}{2}$. The system exhibits relatively stable entanglement generation with minimal oscillatory behavior, suggesting that perturbative dynamics dominate in this regime. Conversely, the strong coupling scenario ($g_0=5.0$, $g_0/\omega=5$) demonstrates rapid, highly oscillatory entanglement dynamics with distinct interference patterns. The emergence of sharp, localized concurrence maxima at early times, followed by complex beating patterns, indicates strong non-perturbative effects where the system-environment interaction becomes comparable to or exceeds the bare system energy scales. We found that when $\theta = \frac{3\pi}{4} + \frac{k\pi}{2}$ (where $k \in \mathbb{Z}$), the initial system state is a decoherence-free state. When $\theta$ is close to $\frac{\pi}{4} + \frac{k\pi}{2}$ (where $k \in \mathbb{Z}$), the coherence fluctuates significantly with time $t$. The frequency of fluctuation is related to $\sqrt{4n+2}g_0$, while the recovery of the peaks is positively correlated with the decay rate $\Gamma$ and the reciprocal of the memory time parameter $\gamma$.

We define a function to measure the entanglement increase from the beginning to the long-time limit as
\scriptsize{
\begin{align}
\Delta(\rho) =& C(\rho)|_{t=\infty} -C(\rho)|_{t=0}\nonumber\\
                        = &\max \{\left|\frac{3}{2} \cos^2 \left(\theta - \frac{\pi}{4}\right)- 1\right| -\frac{\sqrt{n\left(n+1\right)}}{2n+1}\cos^2 \left(\theta - \frac{\pi}{4}\right), 0 \}\nonumber\\
&-\max \{\left|2\cos^2 \left(\theta - \frac{\pi}{4}\right)- 1\right|, 0 \}.
\end{align}
}
\normalsize

\begin{figure}
    \centering
    \includegraphics[width=1.0\columnwidth, clip=true]{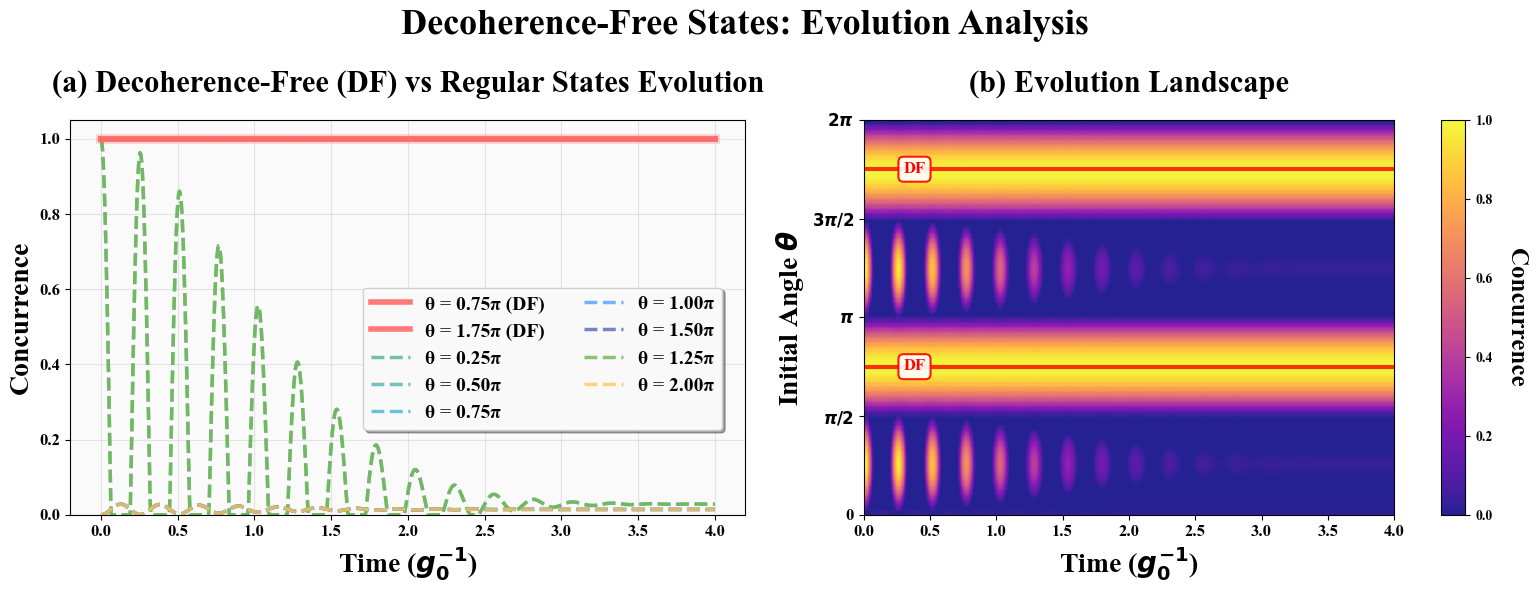}
    \vspace{0mm}
    \caption{Decoherence-free (DF) state analysis: (a) comparison of DF and regular state evolution under strong coupling regime ($g_0=5$, $g_0/\omega=5$); (b) evolution landscape showing DF state locations.}
    \vspace{0mm}
    \label{fig:Fock_separable_state_decoherence_free_analysis}
\end{figure}

Figure~\ref{fig:Fock_separable_state_long_time_entanglement_increase} provides crucial insights into the angular dependence and photon number scaling of entanglement generation efficiency. The periodic structure in Fig.~\ref{fig:Fock_separable_state_long_time_entanglement_increase}(a)(b) with maxima at $\theta = 0, \frac{\pi}{2}, \pi, \frac{3\pi}{2}, 2\pi$ and minima at $\theta = \frac{\pi}{4}, \frac{5\pi}{4}$ reflects the underlying symmetries of the quantum system and suggests that optimal entanglement generation occurs at specific angular orientations with four-fold rotational symmetry. The dramatic decrease in both maximum and minimum entanglement changes with increasing photon number $n$ as shown in Fig.~\ref{fig:Fock_separable_state_long_time_entanglement_increase}(c) demonstrates the classical limit behavior, where quantum correlations become increasingly suppressed as the system approaches macroscopic scales.

Furthermore, the decoherence-free state analysis in Fig.~\ref{fig:Fock_separable_state_decoherence_free_analysis} highlights a remarkable phenomenon where specific quantum states maintain perfect coherence despite environmental coupling. Fig.~\ref{fig:Fock_separable_state_decoherence_free_analysis}(a) demonstrates that decoherence-free states at $\theta = \frac{3\pi}{4}$ and $\theta = \frac{7\pi}{4}$ maintain unit concurrence throughout the entire evolution, contrasting sharply with regular states that rapidly decay to near-zero entanglement.

%%=============================================================%%
%% [R1-3] Added explicit Bell-state identification and dark-state
%% explanation of the decoherence-free mechanism.
%%=============================================================%%
To make the initial-state geometry explicit, we note that at
$\theta = \frac{\pi}{4} + \frac{k\pi}{2}$ ($k = 0, 1, 2, \ldots$) the qubit
pair is prepared in one of the four Bell states (up to a global sign).
Denoting even integers $k = 2j$ and odd integers $k = 2j+1$ ($j = 0,1,2,\ldots$):
\begin{align}
k = 2j: \quad &
  |\psi\rangle = \frac{(-1)^j}{\sqrt{2}}(|01\rangle+|10\rangle)\otimes|n\rangle,
  \label{eq:bell_even}\\
k = 2j+1: \quad &
  |\psi\rangle = \frac{(-1)^{j+1}}{\sqrt{2}}(|01\rangle-|10\rangle)\otimes|n\rangle.
  \label{eq:bell_odd}
\end{align}
The antisymmetric Bell state
$|\psi^-\rangle = \frac{1}{\sqrt{2}}(|01\rangle - |10\rangle)$ is a
\emph{dark state} of the interaction Hamiltonian: one can verify directly that
$H_{\mathrm{tot}}^I|\psi^-,n\rangle = 0$ for all $n$, because the two qubit
contributions $\sigma_-^A|\psi^-\rangle = -\frac{1}{\sqrt{2}}|00\rangle$ and
$\sigma_-^B|\psi^-\rangle = +\frac{1}{\sqrt{2}}|00\rangle$ cancel by
antisymmetry, and similarly for the raising terms.
Consequently, $e^{-i\int_0^t H_{\mathrm{tot}}^I\,d\tau}|\psi^-,n\rangle
= |\psi^-,n\rangle$ exactly for \emph{every} noise realization, so the
concurrence remains unity for all time without any noise-averaging.
This is the microscopic origin of the flat unit-concurrence lines at
$\theta = \frac{3\pi}{4}$ and $\theta = \frac{7\pi}{4}$ in
Fig.~\ref{fig:Fock_separable_state_decoherence_free_analysis}.
For reference, $\theta = \frac{\pi}{2}$ and $\theta = \pi$ correspond to the
unentangled product states $|10\rangle\otimes|n\rangle$ and
$-|01\rangle\otimes|n\rangle$, respectively.

\begin{figure}
    \centering
    \includegraphics[width=1.0\columnwidth, clip=true]{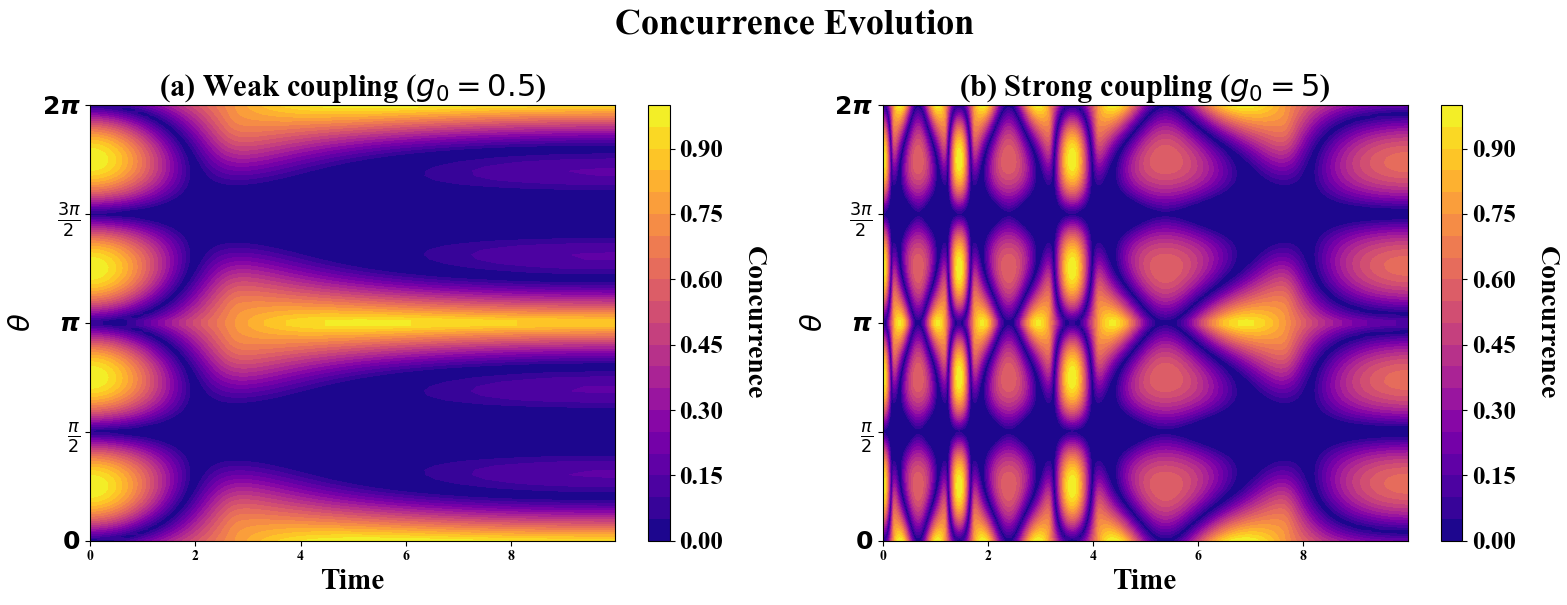}
    \vspace{3mm}
    \caption{Entanglement evolution for initial states $(\cos\theta|00\rangle+\sin\theta|11\rangle) \otimes |1\rangle$ under OU noise ($\Gamma=1$, $\gamma=0.1$): (a) weak coupling regime ($g_0=0.5$, $g_0/\omega=0.5$); (b) strong coupling regime ($g_0=5$, $g_0/\omega=5$).}
    \vspace{-2mm}
    \label{fig:Fock_entangled_state_concurrence}
\end{figure}

\subsection{Case II: Initial State $\cos\theta\,|00\rangle + \sin\theta\,|11\rangle$}

For completeness, we also consider initially entangled qubits:
\begin{equation}
\begin{aligned}
|\psi\rangle=\left(\cos\theta|00\rangle+\sin\theta|11\rangle \right) \otimes |n\rangle.
\end{aligned}
\end{equation}

%%=============================================================%%
%% [R3-e] Added explanatory sentence before excitation sectors.
%%=============================================================%%
Because $|00\rangle$ carries $n$ total excitations while $|11\rangle$ carries
$n+2$, the dynamics separate into two independent excitation sectors for each
photon number $n$.

For a given photon number $n$, the relevant basis states separate into two excitation sectors. The $n$-excitation sector consists of four states:
\begin{small}
\begin{align}
|1\rangle_{n} &= |0,0,n\rangle, \label{eq:basis_n_1}\\
|2\rangle_{n} &= |0,1,n-1\rangle, \label{eq:basis_n_2}\\
|3\rangle_{n} &= |1,0,n-1\rangle, \label{eq:basis_n_3}\\
|4\rangle_{n} &= |1,1,n-2\rangle. \label{eq:basis_n_4}
\end{align}
\end{small}
Similarly, the $(n+2)$-excitation sector contains the corresponding states shifted by two photon quanta:
\begin{small}
\begin{align}
|1\rangle_{n+2} &= |0,0,n+2\rangle, \label{eq:basis_n2_1}\\
|2\rangle_{n+2} &= |0,1,n+1\rangle, \label{eq:basis_n2_2}\\
|3\rangle_{n+2} &= |1,0,n+1\rangle, \label{eq:basis_n2_3}\\
|4\rangle_{n+2} &= |1,1,n\rangle. \label{eq:basis_n2_4}
\end{align}
\end{small}

The Hamiltonian in this basis has a block-diagonal structure:
\begin{equation}
H_{\mathrm{tot}}^{I}(t) = \begin{pmatrix}
H_{n}(t) & 0 \\
0 & H_{n+2}(t)
\end{pmatrix},
\end{equation}
with
\footnotesize{
\begin{equation}
H_{n}\left(t\right)=g\left(t\right) \left(\begin{array}{cccc}
0 & \sqrt{n} & \sqrt{n} & 0\\
\sqrt{n} & 0 & 0 & \sqrt{n-1}\\
\sqrt{n} & 0 & 0 & \sqrt{n-1}\\
0 & \sqrt{n-1} & \sqrt{n-1} & 0 \nonumber
\end{array}\right).
\end{equation}}
\normalsize
After partial tracing the cavity mode and taking the statistical mean over the noise, the reduced density matrix for the qubit system only is
\begin{small}
\begin{equation}
\rho(t) = \mathcal{M}\left[\begin{pmatrix}
Y_{11}^{(n)} & 0 & 0 & Y_{14}^{(n)} \\
0 & Y_{22}^{(n)} & Y_{23}^{(n)} & 0 \\
0 & Y_{23}^{(n)} & Y_{33}^{(n)} & 0 \\
Y_{14}^{(n)} & 0 & 0 & Y_{44}^{(n)}
\end{pmatrix}\right],
\label{eq:rho_general}
\end{equation}
\end{small}

\scriptsize{
\begin{align}
Y_{11}^{(n)} &= \frac{\cos^{2}\theta\left[n - (n-1)\cos x_{n-1}\right]^{2}}{(2n-1)^{2}} \nonumber \\
&\quad + \frac{\sin^{2}\theta\left[\cos x_{n+1} - 1\right]^{2}(n+1)(n+2)}{(2n+3)^{2}} \label{eq:Y11_final}, \\
Y_{14}^{(n)} &= \frac{\cos\theta\sin\theta\left[n - (n-1)\cos x_{n-1}\right]\left[n+2 + (n+1)\cos x_{n+1}\right]}{(2n-1)(2n+3)}
\label{eq:Y14_final}, \\
Y_{22}^{(n)} &= Y_{33}^{(n)} = \frac{n\cos^{2}\theta\sin^{2} x_{n-1}}{4n-2} + \frac{(n+1)\sin^{2}\theta\sin^{2} x_{n+1}}{4n+6}
\label{eq:Y22_final}, \\
Y_{44}^{(n)} &= \frac{\cos^{2}\theta\left[\cos x_{n-1} - 1\right]^{2} n(n-1)}{(2n-1)^{2}} \nonumber \\
&\quad + \frac{\sin^{2}\theta\left[n+2 + (n+1)\cos x_{n+1}\right]^{2}}{(2n+3)^{2}}.
\label{eq:Y44_final}
\end{align}
} \normalsize

Using Eq.~\eqref{eq:phase_accumulation}, $x_m = \sqrt{2(2k+1)}\Omega(t)$ with $\Omega(t) = \int_0^t g(\tau)d\tau$ for $k = n-1$ or $n+1$ as appropriate. The noise averages of these matrix elements are evaluated using the same Gaussian moment identity~\eqref{eq:gaussian_moment} applied to compute $A_n$.

The concurrence is given by
\begin{small}
\begin{align}
C(\rho) &= 2\max\left\{|Y_{14}^{(n)}| - Y_{22}^{(n)} - \sqrt{Y_{11}^{(n)}Y_{44}^{(n)}}, \, 0\right\} \nonumber \\
&= 2\max\Bigg\{\left|\frac{\cos\theta\sin\theta[n - (n-1)\cos x_{n-1}][n+2 + (n+1)\cos x_{n+1}]}{(2n-1)(2n+3)}\right| \nonumber \\
&\quad - \left(\frac{n\cos^{2}\theta\sin^{2} x_{n-1}}{4n-2} + \frac{(n+1)\sin^{2}\theta\sin^{2} x_{n+1}}{4n+6}\right) \nonumber \\
&\quad - \sqrt{\left[\frac{\cos^{2}\theta[n - (n-1)\cos x_{n-1}]^{2}}{(2n-1)^{2}} + \frac{\sin^{2}\theta[\cos x_{n+1} - 1]^{2}(n+1)(n+2)}{(2n+3)^{2}}\right]} \nonumber \\
&\quad \quad \times \sqrt{\left[\frac{\cos^{2}\theta[\cos x_{n-1} - 1]^{2} n(n-1)}{(2n-1)^{2}} + \frac{\sin^{2}\theta[n+2 + (n+1)\cos x_{n+1}]^{2}}{(2n+3)^{2}}\right]}, \, 0\Bigg\}
\label{eq:fock2}
\end{align}
\end{small}

In the long-time limit, the concurrence becomes
\begin{equation}
\Delta(\rho) = C(\rho)|_{t=\infty} -C(\rho)|_{t=0} = -|\sin(2\theta)|
\label{eq:delta_final}
\end{equation}

For the maximally entangled initial state $\theta = \pi/4$, we have $\Delta(\rho)|_{\theta=\pi/4} = -1$. This shows complete loss of entanglement: the system evolves from maximum entanglement ($C=1$) to zero entanglement ($C=0$) in the long-time average. The negative value of $\Delta(\rho)$ indicates entanglement degradation due to dephasing from the cavity interaction, time-averaging over Rabi oscillations, loss of quantum coherence ($Y_{14}^{(n)} \to 0$), and buildup of population in single-excitation states ($Y_{22}^{(n)} > 0$). The magnitude $|\Delta(\rho)| = |\sin(2\theta)|$ is maximized for $\theta = \pi/4$, corresponding to the Bell state $|\Phi^+\rangle = (|00\rangle + |11\rangle)/\sqrt{2}$, which experiences the maximum possible entanglement loss.

Figure~\ref{fig:Fock_entangled_state_concurrence} presents the temporal evolution of quantum entanglement as quantified by concurrence, revealing fundamentally different dynamical regimes as $g_0$ is varied. The contrasting dynamics between initial states $|\psi\rangle=\left(\cos\theta|01\rangle+\sin\theta|10\rangle \right) \otimes |n\rangle$ and $|\psi\rangle=\left(\cos\theta|00\rangle+\sin\theta|11\rangle \right) \otimes |n\rangle$ reveal fundamental aspects of cavity-mediated quantum correlations under noisy conditions. The initial states in the single-excitation subspace demonstrate remarkable entanglement generation capabilities, particularly in low-photon cavity states and near specific angles where qubit configurations optimally couple to the cavity field. In contrast, initially entangled states in the double-excitation subspace exhibit inevitable decoherence without protective configurations. This asymmetry suggests that maintaining quantum correlations requires careful initial state engineering rather than relying solely on dynamical protection.

\section{COHERENT STATE CAVITY}

\subsection{Case I: Initial State $\cos\theta\,|01\rangle + \sin\theta\,|10\rangle$}

One might wonder if the entanglement increase also happens with a coherent state cavity. Let's consider the initial total state in the form
\begin{equation}
\begin{aligned}
|\psi\rangle=\left(\cos\theta|01\rangle+\sin\theta|10\rangle \right) \otimes |\alpha\rangle,
\end{aligned}
\end{equation}
with a coherent state cavity as
\begin{equation}
|\alpha\rangle=e^{-\frac{|\alpha|^{2}}{2}}\sum_{n}\frac{\alpha^{n}}{\sqrt{n!}}|n\rangle=\sum_{n}c_{n}|n\rangle.
\end{equation}

\begin{figure}
\centering
\includegraphics[width=1\columnwidth, clip=true]{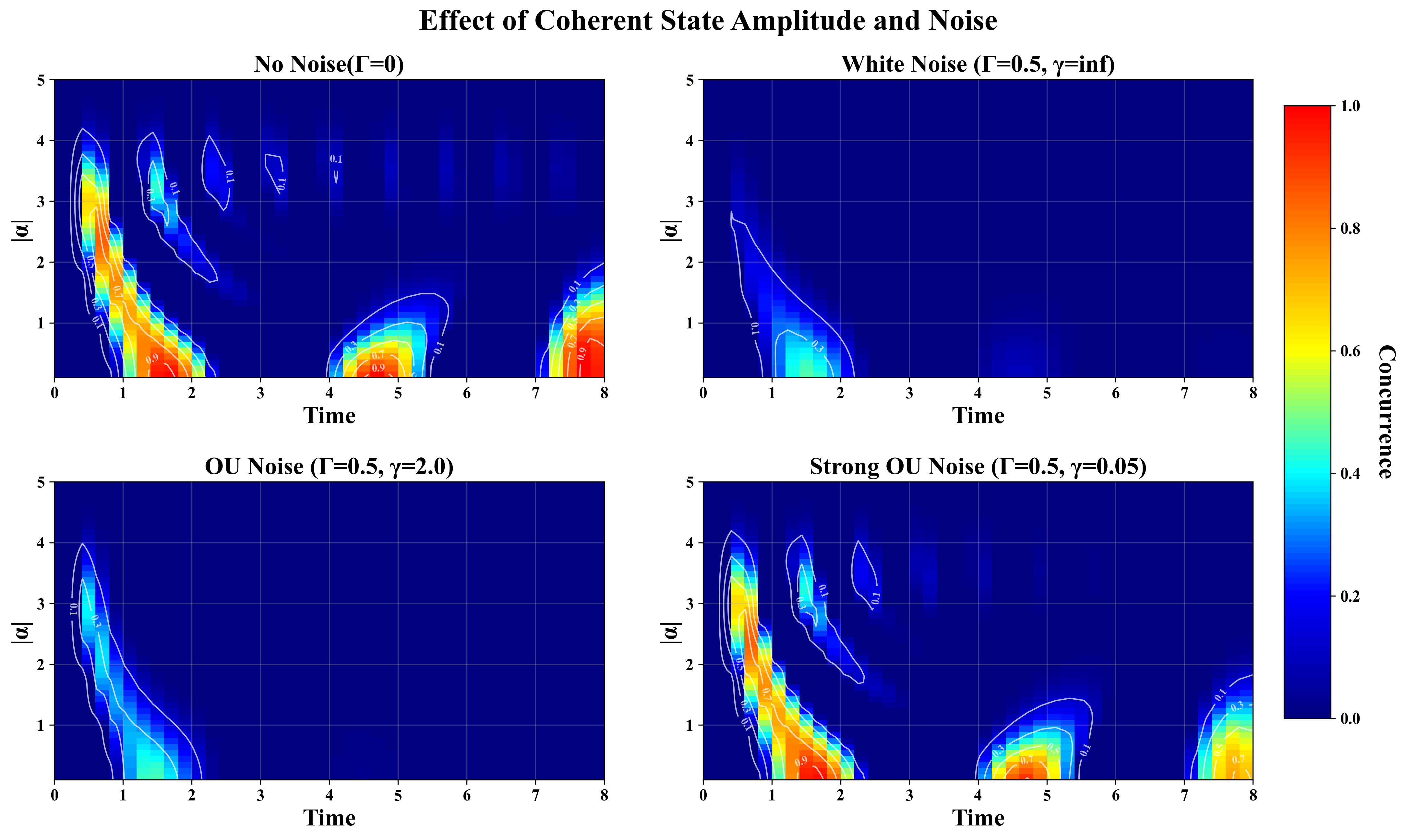}
\caption{Concurrence as a function of time and coherent amplitude $|\alpha|$ for initial state $|\psi\rangle=(\cos\theta|01\rangle+\sin\theta|10\rangle)\otimes|\alpha\rangle$: (a) no noise ($\Gamma=0$); (b) white noise ($\Gamma=0.5$, $\gamma=\infty$); (c) OU noise ($\Gamma=0.5$, $\gamma=2.0$); (d) strong OU noise ($\Gamma=0.5$, $\gamma=0.05$). White contour lines at concurrence levels 0.1, 0.3, 0.5, 0.7, 0.9.}
\label{fig:heatmaps}
\end{figure}

Here, the $|\alpha\rangle$, as the definition of the coherent state, is an eigenstate of the annihilation operator, and the $|n\rangle$ is the Fock state with $n$ photons. Thus, the cavity-qubit system is described as
\begin{equation}
|\psi\rangle=\sum_{n}{c_{n}\left(\cos\theta|01\rangle+\sin\theta|10\rangle \right) \otimes |n\rangle}.
\end{equation}
with coefficient $c_{n} = e^{-\left|\alpha\right|^2}\frac{\alpha^{n}}{\sqrt{n!}}$.
After applying the propagator, obtaining the density matrix for the qubit system alone, and determining the concurrence in the case of this coherent state is straightforward. The reduced-density matrix of the qubit system only is

\begin{eqnarray}
\rho\left(t\right)&=&\sum_{n}\left|c_{n}\right|^2
\mathcal{M}\left[\left(\begin{array}{cccc}
S_{1,n}^{2} & 0 & 0 & 0\\
0 & X_{1,n}^{2} & X_{1,n}X_{2,n} & 0\\
0 & X_{1,n}X_{2,n} & X_{2,n}^{2} & 0\\
0 & 0 & 0 & S_{2,n}^{2}
\end{array}\right)\right].
\end{eqnarray}
where $n$ runs from $0$ to $\infty$ and we use $S_{2,0}=0$. The noise averages of all matrix elements are evaluated using the Gaussian moment identity~\eqref{eq:gaussian_moment} and Eq.~\eqref{eq:An}, as described in Section~3.1.

The coefficients are
\begin{align}
X_{1,n}
 & = -\frac{\sqrt{2}}{2}\sin\!\left(\theta-\frac{\pi}{4}\right)
     + \frac{\sqrt{2}}{2}\cos x_{n}\cos\!\left(\theta-\frac{\pi}{4}\right), \\
X_{2,n}
 & = \frac{\sqrt{2}}{2}\sin\!\left(\theta-\frac{\pi}{4}\right)
     + \frac{\sqrt{2}}{2}\cos x_{n}\cos\!\left(\theta-\frac{\pi}{4}\right), \\
S_{1,n}
 & = \sqrt{\frac{n+1}{2n+1}}\,\sin x_{n}\cos\!\left(\theta-\frac{\pi}{4}\right), \\
S_{2,n}
 & = \sqrt{\frac{n}{2n+1}}\,\sin x_{n}\cos\!\left(\theta-\frac{\pi}{4}\right).
\end{align}
\normalsize

Substituting the matrix elements from $\rho(t)$, we arrive at
\small{
\begin{align}
C(\rho)
 & = 2\max\left\{
     \left|\sum_{n} |c_{n}|^{2}\,\mathcal{M}\!\left[X_{1,n}X_{2,n}\right]\right|
     - \sqrt{
       \sum_{n} |c_{n}|^{2}\mathcal{M}\!\left[S_{1,n}^{2}\right]\,
       \sum_{n} |c_{n}|^{2}\mathcal{M}\!\left[S_{2,n}^{2}\right]
     },
     0
     \right\}.
\end{align}
}

Applying Eq.~\eqref{eq:An} and using $1-A_{n} = \mathcal{M}[\sin^{2}x_{n}]$, we obtain
\begin{small}
\begin{align}
\mathcal{M}\!\left[S_{1,n}^{2}\right]
 & = \frac{n+1}{2n+1}\,
     \cos^{2}\!\left(\theta-\frac{\pi}{4}\right)\,
     \left(1-A_{n}\right), \\
\mathcal{M}\!\left[S_{2,n}^{2}\right]
 & = \frac{n}{2n+1}\,
     \cos^{2}\!\left(\theta-\frac{\pi}{4}\right)\,
     \left(1-A_{n}\right),
\end{align}
\end{small}
with the long time limit of entanglement increase
\small{
\begin{align}
\Delta(\rho) =& C(\rho)|_{t=\infty} -C(\rho)|_{t=0}\nonumber \\
                        =& \max \left\{\left|\frac{3}{2} \cos^2 \left(\theta - \frac{\pi}{4}\right)- 1\right| \nonumber-\sqrt{\sum_{n,m}{\left|c_{n}\right|^2\left|c_{m}\right|^2  \frac{\left(n+1\right)m}{\left(2n+1\right)\left(2m+1\right)}}}\,\cos^2 \left(\theta - \frac{\pi}{4}\right), 0 \right\}\nonumber \\
                        &-\max \left\{\left|2\cos^2 \left(\theta - \frac{\pi}{4}\right)- 1\right|, 0 \right\}.
\end{align}
}
\normalsize

\begin{figure}
\centering
\includegraphics[width=0.8\columnwidth, clip=true]{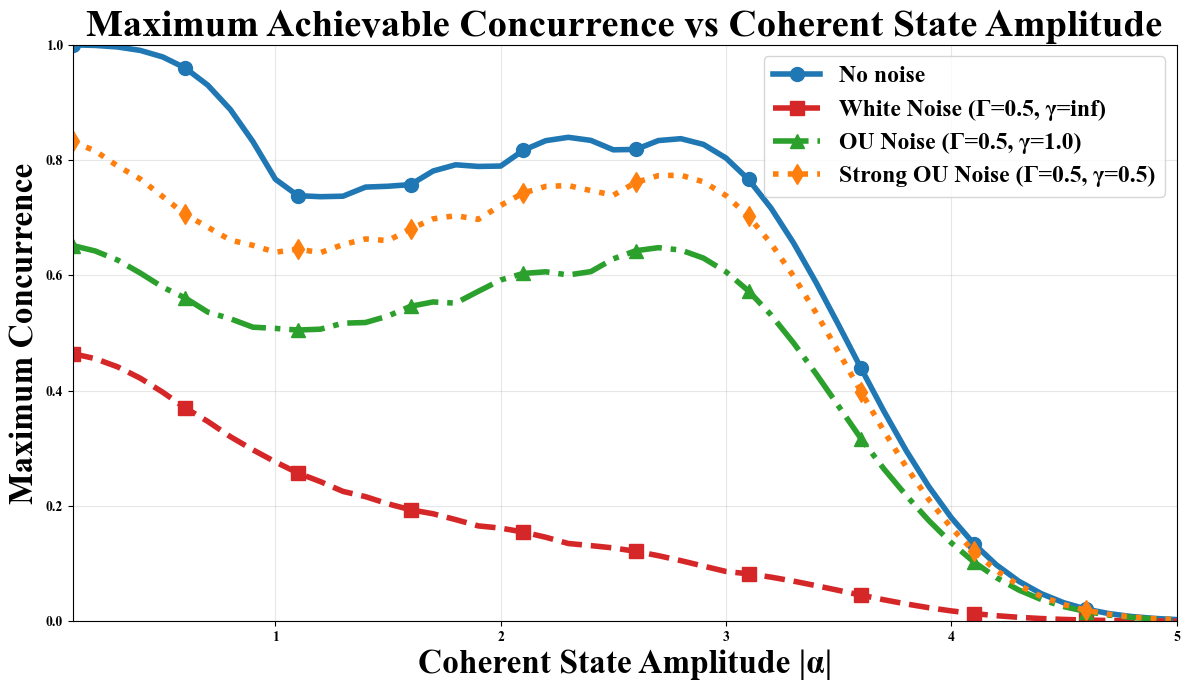}
\vspace{2mm}
\caption{Maximum achievable concurrence versus coherent amplitude $|\alpha|$ for different noise scenarios, optimized over time.}
\vspace{-3mm}
\label{fig:max_concurrence}
\end{figure}

The temporal evolution of entanglement in our stochastic JC model reveals several important physical phenomena as demonstrated in Figs.~\ref{fig:heatmaps} and \ref{fig:max_concurrence}. In the absence of noise, the system exhibits characteristic revival patterns with pronounced periodic structures reflecting the discrete energy level spacing inherent in the cavity-qubit interaction. White noise conditions show rapid and irreversible entanglement decay across the entire parameter space. The monotonic decay observed reflects the Markovian nature of white noise, which introduces uniform decoherence across all time scales. OU noise with moderate correlation time ($\Gamma = 0.5$, $\gamma = 2.0$) presents an intermediate scenario. Most remarkably, strong OU noise with long correlation time ($\Gamma = 0.5$, $\gamma = 0.05$) exhibits behavior qualitatively similar to the noiseless case, with well-defined revival patterns restored, demonstrating that slower environmental fluctuations can assist entanglement preservation by allowing system-environment correlations to persist long enough for information backflow to occur.

Figure~\ref{fig:max_concurrence} quantifies the maximum achievable concurrence as a function of coherent state amplitude across different noise scenarios. In the low-amplitude regime ($|\alpha| < 1$), near-perfect entanglement is achievable in the absence of noise, reflecting the dominance of vacuum and single-photon components. The intermediate amplitude regime exhibits the most striking phenomenon of entanglement revival, with pronounced revival peaks under OU noise conditions. Notably, white noise shows monotonic decay with no revival features.

%%=============================================================%%
%% [R2-4] Expanded explanation of large-|alpha| entanglement death
%% and confirmed white noise has no revivals.
%%=============================================================%%
In the high-amplitude regime ($|\alpha| > 4$, $\langle n \rangle > 16$),
universal entanglement death occurs across all noise conditions.
This behavior has two complementary explanations.
Dynamically, a large-amplitude coherent state involves a broad superposition
of Fock components $|n\rangle$, each evolving at an effective frequency
$\propto\sqrt{4n+2}\,g_0$.
After noise averaging, these incommensurate oscillations interfere
destructively, washing out all coherence in the concurrence formula.
Physically, this reflects the quantum-to-classical transition of the cavity
field: for $|\alpha| \gg 1$ the absolute spread $\Delta n = |\alpha|$ of the
photon-number distribution grows, sampling an increasingly wide range of
coupling frequencies, in close analogy with inhomogeneous broadening.
In the white-noise limit, revivals are absent for \emph{all} $|\alpha|$ because
the decoherence factor $\exp\{-2(4n+2)\Gamma t\}$ decays monotonically for
every $n$; there is no mechanism for information backflow, confirming that
entanglement revivals are a purely non-Markovian phenomenon in our model.

\subsection{Case II: Initial State $\cos\theta\,|00\rangle + \sin\theta\,|11\rangle$}

We begin from the initial state
\begin{align}
|\psi\rangle
 & = \left(\cos\theta\,|00\rangle + \sin\theta\,|11\rangle\right)\otimes|\alpha\rangle \nonumber \\
 & = \sum_{n} c_{n}\left(\cos\theta\,|00\rangle + \sin\theta\,|11\rangle\right)\otimes|n\rangle,
\end{align}
where $|\alpha\rangle = \sum_n c_n |n\rangle = e^{-|\alpha|^2/2}\sum_n \frac{\alpha^n}{\sqrt{n!}}|n\rangle$ is the cavity coherent state.

%%=============================================================%%
%% [R3-e] Added explanatory sentence before excitation sectors.
%%=============================================================%%
As in Section~3.2, because $|00\rangle$ carries $n$ total excitations while
$|11\rangle$ carries $n+2$, the dynamics separate into an $n$-excitation
sector (anchored at $|00,n\rangle$) and an $(n+2)$-excitation sector
(anchored at $|11,n\rangle$) for each photon number $n$.
The relevant basis states in each sector are:

\begin{footnotesize}
\begin{align}
|1\rangle_{n} &= |0,0,n\rangle, \label{eq:coh_basis_n_1}\\
|2\rangle_{n} &= |0,1,n-1\rangle, \label{eq:coh_basis_n_2}\\
|3\rangle_{n} &= |1,0,n-1\rangle, \label{eq:coh_basis_n_3}\\
|4\rangle_{n} &= |1,1,n-2\rangle. \label{eq:coh_basis_n_4}
\end{align}
\end{footnotesize}

Similarly, the $(n+2)$-excitation sector contains:
\begin{footnotesize}
\begin{align}
|1\rangle_{n+2} &= |0,0,n+2\rangle, \label{eq:coh_basis_n2_1}\\
|2\rangle_{n+2} &= |0,1,n+1\rangle, \label{eq:coh_basis_n2_2}\\
|3\rangle_{n+2} &= |1,0,n+1\rangle, \label{eq:coh_basis_n2_3}\\
|4\rangle_{n+2} &= |1,1,n\rangle. \label{eq:coh_basis_n2_4}
\end{align}
\end{footnotesize}

The Hamiltonian in this basis has a block-diagonal structure:
\begin{equation}
H_{\mathrm{tot}}^{I}(t) = \begin{pmatrix}
H_{n}(t) & 0 \\
0 & H_{n+2}(t)
\end{pmatrix}.
\end{equation}

For the coherent state cavity, the total reduced density matrix is
\small{
\begin{equation}
\rho(t) = \sum_{n=0}^{\infty} |c_{n}|^{2}\,
\mathcal{M}\!\left[
\begin{pmatrix}
Y_{11}^{(n)}+Y_{11}^{(n+2)} & 0 & 0 & Y_{14}^{(n,n+2)} \\
0 & Y_{22}^{(n)}+Y_{22}^{(n+2)} & Y_{22}^{(n)}+Y_{22}^{(n+2)} & 0 \\
0 & Y_{22}^{(n)}+Y_{22}^{(n+2)} & Y_{22}^{(n)}+Y_{22}^{(n+2)} & 0 \\
Y_{14}^{(n,n+2)} & 0 & 0 & Y_{44}^{(n)}+Y_{44}^{(n+2)}
\end{pmatrix}
\right].
\end{equation}
}

For this $X$-state, since $\rho_{22} = \rho_{33}$ by symmetry, the concurrence is:
\begin{small}
\begin{align}
C(\rho) & = 2\max\Bigg\{
\left|\sum_{n}|c_{n}|^{2}\mathcal{M}\left[Y_{22}^{(n)}+Y_{22}^{(n+2)}\right]\right| \nonumber \\
&\quad - \sqrt{\sum_{n}|c_{n}|^{2}\mathcal{M}\left[Y_{11}^{(n)}+Y_{11}^{(n+2)}\right] \cdot \sum_{n}|c_{n}|^{2}\mathcal{M}\left[Y_{44}^{(n)}+Y_{44}^{(n+2)}\right]}, \nonumber \\
& \quad \left|\sum_{n}|c_{n}|^{2}\mathcal{M}\left[Y_{14}^{(n,n+2)}\right]\right| - \sum_{n}|c_{n}|^{2}\mathcal{M}\left[Y_{22}^{(n)}+Y_{22}^{(n+2)}\right],\, 0 \Bigg\}.
\end{align}
\end{small}

\begin{figure}[h]
\centering
\includegraphics[width=1\columnwidth, clip=true]{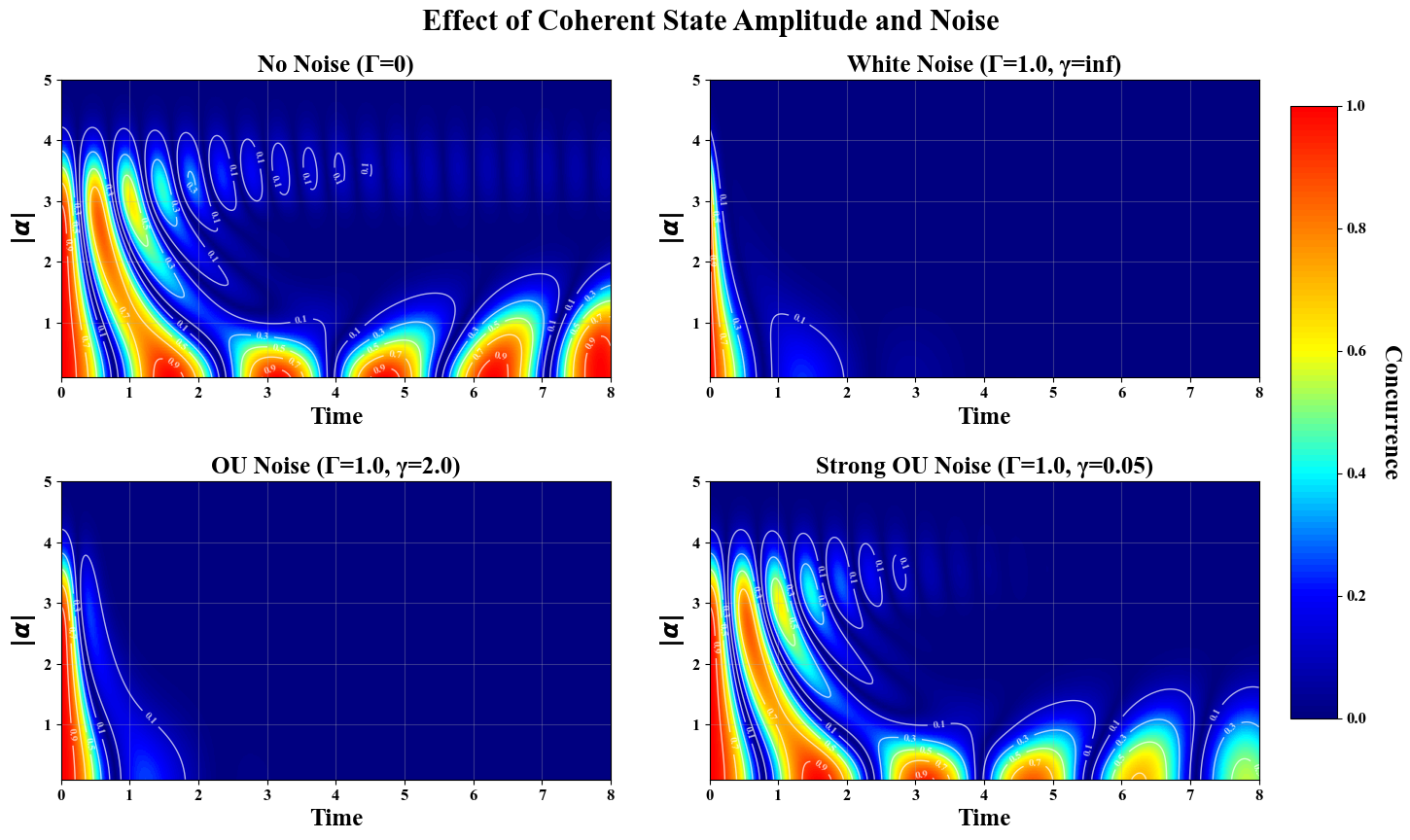}
\caption{Concurrence as a function of time and coherent amplitude $|\alpha|$ for initial state $|\psi\rangle=(\cos\theta|00\rangle+\sin\theta|11\rangle)\otimes|\alpha\rangle$ ($\theta=\pi/4$): (a) no noise ($\Gamma=0$); (b) white noise ($\Gamma=1.0$, $\gamma=\infty$); (c) OU noise ($\Gamma=1.0$, $\gamma=2.0$); (d) strong OU noise ($\Gamma=1.0$, $\gamma=0.05$).}
\label{fig:heatmaps_case2}
\end{figure}

For simplicity, we put the details of this case and the explicit expression of each term in Appendix~\ref{app:coherent2}. The long-time limit of the entanglement increase becomes
\begin{equation}
\Delta(\rho) = C(\rho)|_{t=\infty} - \sin(2\theta),
\end{equation}
where $C(\rho)|_{t=\infty}$ is given by the maximum above with the weighted sums over all photon numbers. Especially, for $\theta = \pi/4$ (equal superposition),
\begin{equation}
\Delta(\rho) = C(\rho)|_{t=\infty} - 1.
\end{equation}
This is always negative (entanglement loss) since the long-time concurrence will be less than the initial maximal entanglement, with the magnitude depending on the coherent state amplitude $|\alpha|$ and noise parameters ($\Gamma$, $\gamma$).

Figure~\ref{fig:heatmaps_case2} presents concurrence heatmaps under four distinct noise conditions. In the absence of noise, the system exhibits rich sudden death and revival dynamics. Under white noise, entanglement undergoes rapid irreversible decoherence with no revival structure, a hallmark of Markovian decoherence. For strong OU noise ($\Gamma = 1.0$, $\gamma = 0.05$), the revival structure partially re-emerges, demonstrating that noise memory rather than noise strength is critical for entanglement survival.

\begin{figure}[h]
\centering
\includegraphics[width=1.0\columnwidth, clip=true]{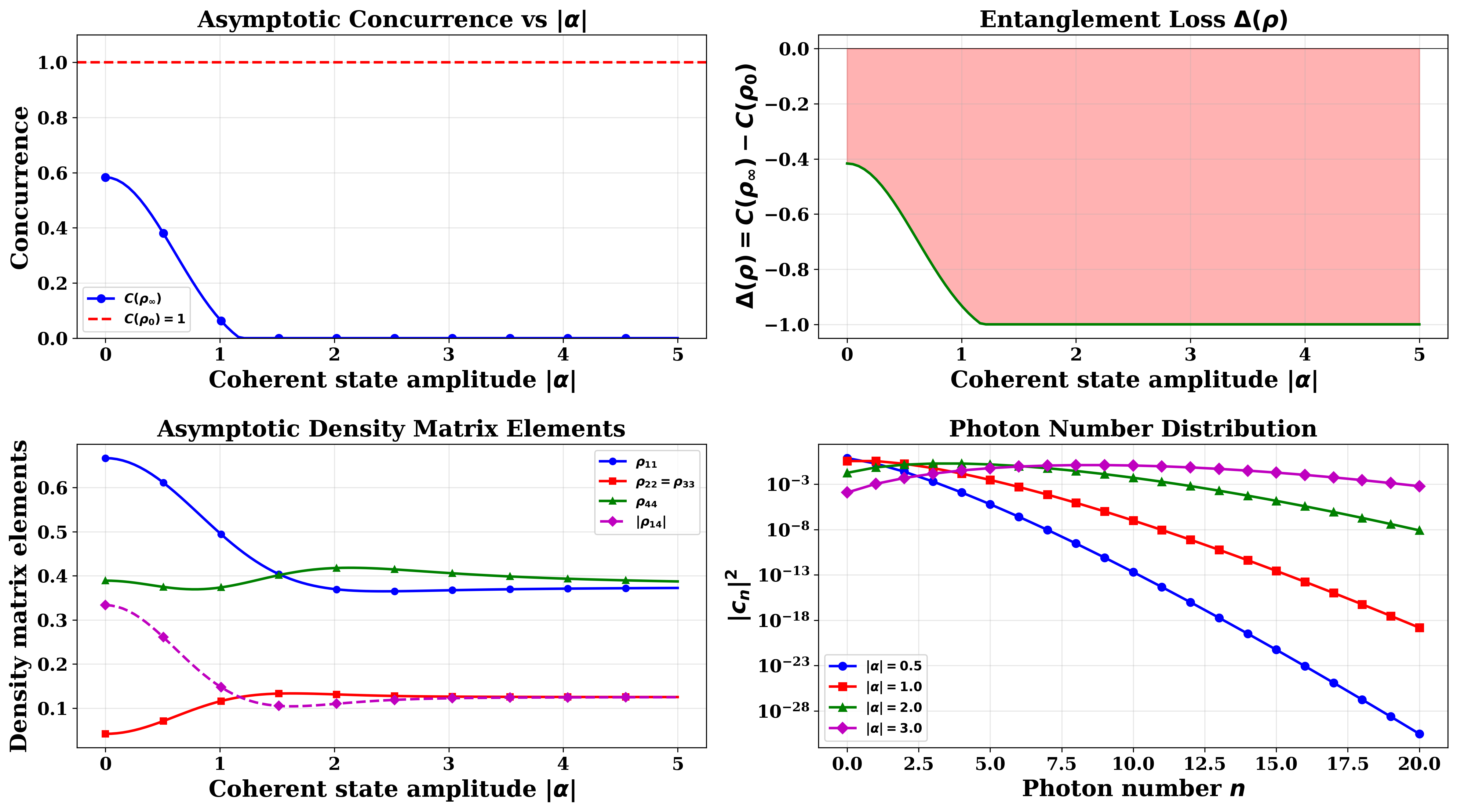}
\caption{Asymptotic entanglement for initial state $\frac{1}{\sqrt{2}}(|00\rangle+|11\rangle)\otimes|\alpha\rangle$: (top left) long-time concurrence $C(\rho_\infty)$ vs.\ $|\alpha|$; (top right) entanglement loss $\Delta(\rho)$; (bottom left) asymptotic density-matrix elements; (bottom right) photon-number distributions $|c_n|^2$ for representative $|\alpha|$.}
\label{fig:coherent_max_2}
\end{figure}

Figure~\ref{fig:coherent_max_2} presents a comprehensive analysis of the long-time entanglement behavior. The asymptotic concurrence $C(\rho_\infty)$ exhibits a dramatic monotonic decrease as a function of $|\alpha|$, starting from $C(\rho_\infty) \approx 0.583$ in the vacuum limit and decaying to near-zero for $|\alpha| \gtrsim 1.5$. The complete loss of entanglement for $|\alpha| \geq 2$ suggests that quantum information protocols relying on entangled qubits in cavity QED architectures must operate in the low-photon regime.

Overall, these results have important implications for quantum information processing applications. We identify three distinct optimal operating regimes: low amplitudes maximize entanglement preservation under noise-free conditions, intermediate amplitudes enable entanglement revival when coupled with engineered non-Markovian environments, while high amplitudes should be avoided in any noisy environment due to universal entanglement death. The fundamental distinction lies in the environmental memory: Markovian processes cause irreversible information loss through memoryless interactions, while non-Markovian processes with extended correlation times enable partial quantum coherence restoration through memory-assisted revival mechanisms.

%%=============================================================%%
%% [R3-g,h / R2-5] Rewrote conclusion to lead with analytical
%% results and added cavity-decay limitation discussion.
%%=============================================================%%
\section{CONCLUSION AND OUTLOOK}

We have analyzed entanglement dynamics in a two-qubit Jaynes-Cummings model
with stochastic coupling $g(t)=g_0+\eta(t)$, where $\eta(t)$ is either
white noise (Markovian) or an Ornstein-Uhlenbeck process (non-Markovian),
for both Fock and coherent cavity initial states.

\paragraph{Main analytical results.}
(i)~We derived closed-form expressions for the noise-averaged concurrence
through the parameter $A_n$ (Eq.~\eqref{eq:An}), which encodes coherent Rabi
oscillations and exponential OU decoherence in a single formula and reduces
exactly to the white-noise result as $\gamma\to\infty$.
(ii)~For the single-excitation initial state
$(\cos\theta|01\rangle+\sin\theta|10\rangle)\otimes|n\rangle$, the
antisymmetric Bell state $|\psi^-\rangle$ is a dark state of
$H_{\mathrm{tot}}^I$ satisfying $H_{\mathrm{tot}}^I|\psi^-,n\rangle=0$,
rendering it exactly decoherence-free for all noise realizations.
(iii)~For initially entangled states
$(\cos\theta|00\rangle+\sin\theta|11\rangle)\otimes|n\rangle$, the long-time
entanglement loss satisfies $\Delta(\rho) = -|\sin(2\theta)|$, proving that
all pre-existing entanglement is irrecoverably lost; no protective subspace
exists for this class.
(iv)~For coherent cavity states, we identify a photon-number threshold in
the intermediate amplitude regime separating low-frequency from
high-frequency entanglement oscillations, and prove that universal
entanglement death at large amplitudes arises from destructive interference
among incommensurate Fock-component oscillations.

\paragraph{Physical implications.}
Non-Markovian environments with long correlation times can outperform Markovian
noise through memory-assisted revival: the OU decoherence factor recovers
partially between oscillation cycles, whereas white noise decays monotonically
with no information backflow.
These findings establish cavity photon statistics and environmental memory time
as key control parameters for engineering entanglement dynamics.
Optimal operating regimes for entanglement protocols are: low coherent
amplitudes for noise-free operation; intermediate amplitudes with engineered
non-Markovian environments for revival-based protocols; and the vacuum or
few-photon Fock regime for maximum entanglement generation efficiency.

\paragraph{Limitations and future directions.}
Our model assumes an ideal lossless cavity.
In practice, photon leakage at rate $\kappa$ introduces additional decoherence
via the Lindblad dissipator
$\kappa(a\rho a^\dagger - \frac{1}{2}\{a^\dagger a,\rho\})$.
For $\kappa \ll g_0$ the dominant entanglement dynamics remain governed by the
stochastic coupling and our results provide an upper bound on entanglement
lifetime; for $\kappa \sim g_0$ (bad-cavity limit), leakage would suppress
revivals even under OU noise.
Incorporating cavity decay via a stochastic Lindblad master equation,
extending the analysis to squeezed or thermal cavity states, and studying
multi-qubit network topologies are natural directions for future work.
Experimental verification of the predicted revival phenomena in circuit-QED
platforms, where noise correlation times and coupling strengths are tunable,
would provide direct tests of the non-Markovian protection mechanism identified
here.

\backmatter

\begin{appendices}

\section{Special Case for Fock State Cavity: when $n=0$ and $n=1$}
\label{app:note_fock_special_case}

\subsection*{Initial State of (1)}

If we assume the two qubits are coupled to the vacuum, then $n=0$.
For the initial state (1), the basis of Hilbert space is reduced to $|1\rangle = |0,0,1\rangle$, $|2\rangle = |0,1,0\rangle$, $|3\rangle = |1,0,0\rangle$. In the resonant case, the Hamiltonian of the system is reduced to
\begin{equation}
H_{0}(t)=\left(\begin{array}{ccc}
0 & g(t) & g(t)\\
g(t) & 0 & 0\\
g(t) & 0 & 0
\end{array}\right),
\end{equation}
and the propagator becomes
\begin{equation}
G_{0}(t)=\left(\begin{array}{ccc}
\cos x_{0} & -\frac{i\sin x_{0}}{\sqrt{2}} & -\frac{i\sin x_{0}}{\sqrt{2}}\\
\frac{i\sin x_{0}}{\sqrt{2}} & \frac{\cos x_{0}+1}{2} & \frac{\cos x_{0}-1}{2}\\
\frac{i\sin x_{0}}{\sqrt{2}} & \frac{\cos x_{0}-1}{2} & \frac{\cos x_{0}+1}{2}
\end{array}\right),
\end{equation}
where
\begin{equation}
x_{0}=\sqrt{2}\Omega(t)=\sqrt{2}\int_{0}^{t}g(\tau)d\tau.
\end{equation}
Similarly, the density matrix after time evolution becomes a more simplified form
\begin{equation}
\rho\left(t\right)=\mathcal{M}\left[\left(\begin{array}{ccc}
S_{1,n}^{2} & -iS_{1,n}X_{1,n} & -iS_{1,n}X_{2,n}\\
iS_{1,n}X_{1,n} & X_{1,n}^{2} & X_{1,n}X_{2,n}\\
iS_{1,n}X_{2,n} & X_{1,n}X_{2,n} & X_{2,n}^{2}
\end{array}\right)\right],
\end{equation}
where
\begin{align}
    S_{1,n}&=\frac{\sin x_{0}(\cos\theta+\sin\theta)}{\sqrt{2}}, \\
    X_{1,n}&=\frac{1}{2}[(\cos x_{0}+1)\cos\theta+(\cos x_{0}-1)\sin\theta], \\
    X_{2,n}&=\frac{1}{2}[(\cos x_{0}-1)\cos\theta+(\cos x_{0}+1)\sin\theta].
\end{align}
To obtain the concurrence, the cavity part needs to be traced out. Calculating the partial trace of the density matrix, the reduced density matrix of the qubits is as follows:
\begin{equation}
\mathcal{M}[\rho_{\psi}^{Q}(t)]=\begin{pmatrix}\mathcal{M}[S_{1,n}^{2}] & 0 & 0 & 0\\
0 & \mathcal{M}[X_{1,n}^{2}] & \mathcal{M}[X_{1,n}X_{2,n}] & 0\\
0 & \mathcal{M}[X_{1,n}X_{2,n}] & \mathcal{M}[X_{2,n}^{2}] & 0\\
0 & 0 & 0 & 0
\end{pmatrix}.
\end{equation}
From the definition, the concurrence is
\begin{align}
C_{\rho}(t) & = 2\max\{|\mathcal{M}[X_{1,n}X_{2,n}]|,0\} \nonumber \\
& = \max \{\left|\left(A+1\right) \cos^2 \left(\theta - \frac{\pi}{4}\right)- 1\right|, 0 \},
\end{align}
with
\begin{equation}
A=\exp\left\{ -4\left[\Gamma t +\frac{\Gamma}{\gamma}(e^{-\gamma t}-1)\right]  \right\}.
\end{equation}
The asymptotic concurrence when $t\rightarrow\infty$ is
\begin{equation}
C_{\rho}(t\rightarrow\infty) = \max \{\left|\frac{3}{2}\cos^2 \left(\theta - \frac{\pi}{4}\right)- 1\right|,0\}.
\end{equation}

\subsection*{Initial State of (2)}
Considering $n=0$, the qubits are coupled with a single mode with one photon.
For initial state (2), the basis of Hilbert space is $|1\rangle_{1} = |0,0,1\rangle$, $|2\rangle_{1} = |0,1,0\rangle$, $|3\rangle_{1} = |1,0,0\rangle$, $|1\rangle_{2} = |0,0,3\rangle$, $|2\rangle_{2} = |0,1,2\rangle$, $|3\rangle_{2} = |1,0,2\rangle$, $|4\rangle_{2} = |1,1,1\rangle$.

In the resonant case, the Hamiltonian of the system is reduced to
\begin{equation}
H_{0,2}(t) = \left(\begin{array}{cc}
H_{0}(t) & 0\\
0 & H_{2}(t)
\end{array}\right).
\end{equation}
The time evolution operator is
\begin{equation}
U_{0,2}(t) = \left(\begin{array}{cc}
U_{0}(t) & 0\\
0 & U_{2}(t)
\end{array}\right).
\end{equation}
After tracing out the cavity mode, we have
\begin{equation}
\rho_{\phi}^{Q}(t) = \left(\begin{array}{cccc}
Y_{11} & 0 & 0 & e^{2i\omega t}Y_{14}\\
0 & Y_{22} & Y_{23} & 0\\
0 & Y_{32} & Y_{33} & 0\\
e^{-2i\omega t}Y_{14} & 0 & 0 & Y_{44}
\end{array}\right),
\end{equation}
where
\begin{align}
    Y_{11} &= \cos^{2}\theta\cos^{2}x_{0} + \frac{6\sin^{2}\theta(\cos X_{2,n}-1)^{2}}{25}, \\
    Y_{14} &= \frac{\cos\theta\sin\theta\cos x_{0}(3+2\cos X_{2,n})}{5}, \\
    Y_{22} &= \frac{\cos^{2}\theta\sin^{2}x_{0}}{2} + \frac{2\sin^{2}\theta\sin^{2}X_{2,n}}{10}, \\
    Y_{44} &= \frac{\sin^{2}\theta(3+2\cos X_{2,n})^{2}}{25}.
\end{align}
This is the same result as Eq.~\ref{special1} by taking $n=0$.

When $n=-1$, the qubits are coupled with the vacuum. For the initial state (3), the state can be simplified to $|\phi,0\rangle = \cos\theta|00,0\rangle + \sin\theta|11,0\rangle$.
The first term does not evolve. We only need to consider the second term, for which the basis of Hilbert space is $|1\rangle_{1} = |0,0,2\rangle$, $|2\rangle_{1} = |0,1,1\rangle$, $|3\rangle_{2} = |1,0,1\rangle$, $|4\rangle_{2} = |1,1,0\rangle$. In the resonant case, the Hamiltonian is
\begin{equation}
H_{1}(t) = \begin{pmatrix}
\omega & g(t)\sqrt{2} & g(t)\sqrt{2} & 0\\
g(t)\sqrt{2} & \omega & 0 & g(t)\\
g(t)\sqrt{2} & 0 & \omega & g(t)\\
0 & g(t) & g(t) & \omega
\end{pmatrix}.
\end{equation}

Following similar derivation procedures, the expectation of the reduced density matrix for $|\phi,0\rangle$ becomes
\begin{equation}
\mathcal{M}[\rho_{\phi}^{Q}(t)] = \left(\begin{array}{cccc}
\mathcal{M}[Y_{11}] & 0 & 0 & 0\\
0 & \mathcal{M}[Y_{22}] & \mathcal{M}[Y_{23}] & 0\\
0 & \mathcal{M}[Y_{32}] & \mathcal{M}[Y_{33}] & 0\\
0 & 0 & 0 & \mathcal{M}[Y_{44}]
\end{array}\right),
\end{equation}
where
\begin{align}
Y_{11} &= \cos^{2}\theta + \frac{2\sin^{2}\theta(\cos X_{1,n}-1)^{2}}{9}, \\
Y_{22} &= \frac{\sin^{2}\theta\sin^{2}X_{1,n}}{6}, \\
Y_{44} &= \frac{\sin^{2}\theta(2+\cos X_{1,n})^{2}}{9}.
\end{align}

The asymptotic concurrence is
\begin{align}
C_{\psi}(t\rightarrow\infty) &= 2\max\left\{\frac{\sin^{2}\theta}{12}-\sqrt{\left(\cos^{2}\theta+\frac{\sin^{2}\theta}{3}\right)\frac{7\sin^{2}\theta}{12}},0\right\} = 0.
\end{align}

\section{Detailed time evolution for $\cos\theta\,|00\rangle + \sin\theta\,|11\rangle$ initial state with coherent cavity}
\label{app:coherent2}

In this appendix, we provide the full derivation of the reduced two-qubit density matrix and concurrence for the initially entangled state coupled with a coherent cavity in Case II of the main text. We begin from the initial state
\begin{align}
|\psi\rangle
 & = \left(\cos\theta\,|00\rangle + \sin\theta\,|11\rangle\right)\otimes|\alpha\rangle \nonumber \\
 & = \sum_{n} c_{n}\left(\cos\theta\,|00\rangle + \sin\theta\,|11\rangle\right)\otimes|n\rangle,
\end{align}
where $|\alpha\rangle = e^{-|\alpha|^2/2}\sum_n \frac{\alpha^n}{\sqrt{n!}}|n\rangle$ is the cavity coherent state with $c_n = e^{-|\alpha|^2/2}\frac{\alpha^n}{\sqrt{n!}}$. The total Hamiltonian in the interaction picture is
\begin{equation}
H_{\mathrm{tot}}^{I}(t) = g(t) \left[ a^{\dagger} \left( \sigma_{-}^{A} + \sigma_{-}^{B} \right) + \text{H.c.} \right],
\label{eq:interaction_hamiltonian_appC}
\end{equation}
with time-evolution operator as Eq.~(\ref{eq:propagator}). The initial state $|00,n\rangle$ has total excitation number $n$, while $|11,n\rangle$ has excitation number $n+2$. The dynamics separate into independent blocks for each photon number. The Hamiltonian has a block-diagonal structure:
\begin{equation}
H_{\mathrm{tot}}^{I}(t) = \begin{pmatrix}
H_{n}(t) & 0 \\
0 & H_{n+2}(t)
\end{pmatrix},
\end{equation}
where
\begin{small}
\begin{equation}
H_{n}(t) = g(t)\,
\begin{pmatrix}
0 & \sqrt{n} & \sqrt{n} & 0 \\
\sqrt{n} & 0 & 0 & \sqrt{n-1} \\
\sqrt{n} & 0 & 0 & \sqrt{n-1} \\
0 & \sqrt{n-1} & \sqrt{n-1} & 0
\end{pmatrix},
\end{equation}
\end{small}
in the basis $\{|00,n\rangle, |01,n-1\rangle, |10,n-1\rangle, |11,n-2\rangle\}$, and similarly for $H_{n+2}(t)$ with $n \to n+2$.

The time-evolution operator $G_{n}(t) = \exp[-i\int_{0}^{t} H_{n}(\tau)\,d\tau]$ can be diagonalized. The key matrix elements are:
\begin{small}
\begin{align}
G_{n,11} &= \frac{n-1+n\cos x_{n-1}}{2n-1}, \label{eq:Gn11}\\
G_{n,21} &= -i\frac{\sqrt{n}\sin x_{n-1}}{\sqrt{4n-2}}, \label{eq:Gn21}\\
G_{n,31} &= -i\frac{\sqrt{n}\sin x_{n-1}}{\sqrt{4n-2}}, \label{eq:Gn31}\\
G_{n,41} &= -\frac{\sqrt{n(n-1)}\left(1-\cos x_{n-1}\right)}{2n-1}. \label{eq:Gn41}
\end{align}
\end{small}
where $x_{n-1} = \sqrt{4n-2}\,\Omega(t)$ with $\Omega(t) = \int_{0}^{t} g(\tau)\,d\tau$. For the $(n+2)$-block, $x_{n+1} = \sqrt{4n+6}\,\Omega(t)$. After tracing out the cavity and applying stochastic averaging $\mathcal{M}[\cdot]$ via the Gaussian moment identity~\eqref{eq:gaussian_moment}, the reduced density matrix has $X$-form:

\begin{equation}
\rho(t) = \sum_{n=0}^{\infty} |c_{n}|^{2}\,
\mathcal{M}\!\left[
\begin{pmatrix}
Y_{11}^{(n)}+Y_{11}^{(n+2)} & 0 & 0 & Y_{14}^{(n,n+2)} \\
0 & Y_{22}^{(n)}+Y_{22}^{(n+2)} & Y_{22}^{(n)}+Y_{22}^{(n+2)} & 0 \\
0 & Y_{22}^{(n)}+Y_{22}^{(n+2)} & Y_{22}^{(n)}+Y_{22}^{(n+2)} & 0 \\
Y_{14}^{(n,n+2)} & 0 & 0 & Y_{44}^{(n)}+Y_{44}^{(n+2)}
\end{pmatrix}
\right],
\end{equation}
where the matrix elements before averaging are:
\begin{small}
\begin{align}
Y_{11}^{(n)} &= \cos^{2}\theta\left|\frac{n-1+n\cos x_{n-1}}{2n-1}\right|^{2}, \label{eq:appY11n}\\
Y_{22}^{(n)} &= \frac{n\cos^{2}\theta\sin^{2}x_{n-1}}{4n-2}, \label{eq:appY22n}\\
Y_{44}^{(n)} &= \frac{\cos^{2}\theta\,n(n-1)(1-\cos x_{n-1})^{2}}{(2n-1)^{2}}, \label{eq:appY44n}
\end{align}
\end{small}
and for the $(n+2)$-sector:
\begin{small}
\begin{align}
Y_{11}^{(n+2)} &= \frac{\sin^{2}\theta(n+1)(n+2)(1-\cos x_{n+1})^{2}}{(2n+3)^{2}}, \label{eq:appY11n2}\\
Y_{22}^{(n+2)} &= \frac{(n+1)\sin^{2}\theta\sin^{2}x_{n+1}}{4n+6}, \label{eq:appY22n2}\\
Y_{44}^{(n+2)} &= \sin^{2}\theta\left|\frac{(n+2)+(n+1)\cos x_{n+1}}{2n+3}\right|^{2}, \label{eq:appY44n2}
\end{align}
\end{small}
with off-diagonal coherence
\begin{align}
Y_{14}^{(n,n+2)} = & \frac{\cos\theta\sin\theta}{(2n-1)(2n+3)} \nonumber \\
& \times (n-1+n\cos x_{n-1}) \nonumber \\
& \times ((n+2)+(n+1)\cos x_{n+1}). \label{eq:appY14}
\end{align}

For the $X$-state, the concurrence is
\begin{equation}
C(\rho) = 2\max\Big\{|\rho_{23}| - \sqrt{\rho_{11}\rho_{44}}, \, |\rho_{14}| - \rho_{22}, \, 0\Big\}.
\end{equation}

Applying Eq.~(\ref{eq:An}), the noise-averaged elements are:
\begin{small}
\begin{align}
\mathcal{M}[Y_{11}^{(n)}] &= \frac{\cos^{2}\theta}{(2n-1)^{2}}
    \Big[(n-1)^2 \nonumber \\
&\quad + 2n(n-1)\mathcal{M}[\cos x_{n-1}] + n^2 A_{n-1}\Big], \\
\mathcal{M}[Y_{22}^{(n)}] &= \frac{n\cos^{2}\theta}{4n-2}(1 - A_{n-1}), \\
\mathcal{M}[Y_{44}^{(n)}] &= \frac{\cos^{2}\theta\,n(n-1)}{(2n-1)^{2}}
    \Big[1 - 2\mathcal{M}[\cos x_{n-1}] + A_{n-1}\Big].
\end{align}
\end{small}
Similar expressions hold for the $(n+2)$-sector with $n \to n+1$ in the subscripts. For the cross-sector term:
\begin{small}
\begin{align}
\mathcal{M}[Y_{14}^{(n,n+2)}] &= \frac{\cos\theta\sin\theta}{(2n-1)(2n+3)}
    \Big[(n-1)(n+2) \nonumber \\
& + (n-1)(n+1)\mathcal{M}[\cos x_{n+1}] \nonumber \\
& + n(n+2)\mathcal{M}[\cos x_{n-1}] \nonumber \\
& + n(n+1)\mathcal{M}[\cos x_{n-1}\cos x_{n+1}]\Big].
\end{align}
\end{small}

In the long-time limit, all oscillatory terms vanish:
\begin{equation}
\mathcal{M}[\cos x_n] \to 0, \quad \mathcal{M}[\cos^2 x_n] \to 1/2.
\end{equation}

The asymptotic matrix elements become:
\begin{small}
\begin{align}
\mathcal{M}[Y_{11}^{(n)}]|_{t\to\infty} &= \frac{\cos^2\theta(3n^2-4n+2)}{2(2n-1)^2}, \\
\mathcal{M}[Y_{22}^{(n)}]|_{t\to\infty} &= \frac{n\cos^2\theta}{4(2n-1)}, \\
\mathcal{M}[Y_{44}^{(n)}]|_{t\to\infty} &= \frac{3\cos^2\theta \cdot n(n-1)}{2(2n-1)^2},
\end{align}
\end{small}
and for the $(n+2)$-sector:
\begin{small}
\begin{align}
\mathcal{M}[Y_{11}^{(n+2)}]|_{t\to\infty} &= \frac{3\sin^2\theta(n+1)(n+2)}{2(2n+3)^2}, \\
\mathcal{M}[Y_{22}^{(n+2)}]|_{t\to\infty} &= \frac{(n+1)\sin^2\theta}{4(2n+3)}, \\
\mathcal{M}[Y_{44}^{(n+2)}]|_{t\to\infty} &= \frac{\sin^2\theta(3n^2+14n+14)}{2(2n+3)^2}, \\
\mathcal{M}[Y_{14}^{(n,n+2)}]|_{t\to\infty} &= \frac{\cos\theta\sin\theta(n-1)(n+2)}{(2n-1)(2n+3)}.
\end{align}
\end{small}
The entanglement change is defined as
\begin{equation}
\Delta(\rho) = C(\rho)|_{t=\infty} - \sin(2\theta),
\end{equation}
where $C(\rho)|_{t=\infty}$ is computed from the asymptotic density matrix. For $\theta = \pi/4$:
\begin{equation}
\Delta(\rho) = C(\rho)|_{t=\infty} - 1,
\end{equation}
which is always negative, indicating entanglement loss. The magnitude depends on the coherent state amplitude $|\alpha|$ and noise parameters ($\Gamma$, $\gamma$).

\section{Scaling of Cavity Frequency and Coupling Fluctuations}
\label{app:Scaling}

In this appendix we provide a simple estimate of how small fluctuations of the cavity size translate into fluctuations of the cavity frequency $\omega_f$ and the atom--cavity coupling strength $g(t)$, and how this connects to the Ornstein-Uhlenbeck (OU) model used in the main text.

\subsubsection{Cavity frequency fluctuations from length changes}

We first consider a one-dimensional Fabry--Perot cavity of length $L$ and refractive index $n$ supporting a standing-wave mode with index $m$. The resonance frequency is
\begin{equation}
\omega_m(L) = \frac{m\pi c}{n L}.
\end{equation}
A small change of the cavity length, $L \rightarrow L + \delta L$ with $|\delta L| \ll L$, induces a corresponding shift of the mode frequency
\begin{equation}
\delta \omega_f \approx \frac{\partial \omega_m}{\partial L}\,\delta L = -\frac{m\pi c}{n L^2}\,\delta L = -\frac{\omega_f}{L}\,\delta L.
\end{equation}
Therefore the relative fluctuation of the cavity frequency is
\begin{equation}
\frac{\delta \omega_f}{\omega_f} \approx -\frac{\delta L}{L}.
\label{eq:freq_scaling}
\end{equation}
If the cavity length change is caused by thermal expansion, we can write $\delta L \approx \alpha_L L\,\Delta T$, where $\alpha_L$ is the linear thermal expansion coefficient and $\Delta T$ is a small temperature excursion around the operating point. Substituting this into Eq.~\eqref{eq:freq_scaling} yields
\begin{equation}
\frac{\delta \omega_f}{\omega_f} \approx -\alpha_L \Delta T.
\end{equation}

\subsubsection{Coupling strength fluctuations from cavity geometry}

The atom--cavity coupling strength can be written as $g = \boldsymbol{\mu}\cdot\mathbf{E}_{\mathrm{vac}}/\hbar$, where $\boldsymbol{\mu}$ is the transition dipole moment and $\mathbf{E}_{\mathrm{vac}}$ is the vacuum electric field amplitude. For a single mode with angular frequency $\omega_f$ and mode volume $V$, the vacuum field scales as $E_{\mathrm{vac}} \propto \sqrt{\hbar \omega_f/(2\varepsilon_0 V)}$, so that $g \propto \sqrt{\omega_f/V}$. For a Fabry--Perot cavity with cross-sectional area $A$ and length $L$, the mode volume is $V \sim A L$, and using $\omega_f \propto 1/L$ we obtain $g(L) \propto 1/L$. A small change in cavity length thus produces
\begin{equation}
\frac{\delta g}{g} \approx -\frac{\delta L}{L} \approx -\alpha_L \Delta T.
\label{eq:g_scaling}
\end{equation}
Equations~\eqref{eq:freq_scaling} and \eqref{eq:g_scaling} show that both the cavity frequency and the coupling strength inherit comparable relative fluctuations from small changes of the cavity length.

\subsubsection{Connection to the Ornstein-Uhlenbeck model}

To connect these geometric considerations to the stochastic model in the main text, we assume that the cavity length fluctuates according to an Ornstein-Uhlenbeck process, $L(t) = L_0 + \delta L(t)$, with zero mean and correlation function $\mathcal{M}[\delta L(t)\,\delta L(s)] = \sigma_L^2 e^{-|t-s|/\tau_L}$. Using the linearized relation $\delta g(t) \approx -(g_0/L_0)\,\delta L(t)$, we obtain
\begin{equation}
\mathcal{M}[\delta g(t)\,\delta g(s)] = \left(\frac{g_0}{L_0}\right)^2 \sigma_L^2 e^{-|t-s|/\tau_L}.
\end{equation}
Identifying $\eta(t) \equiv \delta g(t)$, this correlation function is of the same form as Eq.~\eqref{eq:ou_correlation}, with correspondences
\begin{equation}
\gamma \sim \frac{1}{\tau_L}, \qquad \Gamma \sim 2\left(\frac{g_0}{L_0}\right)^2 \sigma_L^2.
\end{equation}
Thus, an OU-type fluctuation of the cavity length naturally induces OU fluctuations of both $\omega_f(t)$ and $g(t)$ with the same correlation time, establishing a concrete physical realization of the stochastic coupling model $g(t) = g_0 + \eta(t)$ used in the main text.

\end{appendices}

\bmhead{Acknowledgements} No acknowledgments.

\bmhead{Author Contributions} F.W. and Q.Z.D. were responsible for conceptualization, model derivation, and development of the numerical solutions. F.W. conducted the literature review and prepared the manuscript. J.J. and T.Y. provided supervision and contributed to conceptualization. All authors have reviewed and approved the final manuscript.

\bmhead{Data availability} No datasets were generated or analyzed during the current study.

\bmhead{Code Availability} Codes will be shared upon reasonable request.

\section*{Declarations}

\bmhead{Conflict of interest} The authors declare no conflicts of interest.

\bmhead{Funding} This research did not receive any funding.

\bmhead{Ethical approval and consent to participate} Not applicable.

\bibliography{sn-bibliography}

\end{document}